\documentclass[aps]{revtex4-2}
\usepackage{amsmath,amssymb,bbold}
\usepackage{color,graphicx}
\usepackage{geometry}
\newgeometry{tmargin=2cm, bmargin=3cm, lmargin=1cm, rmargin=1cm}
\newcommand{\be}{\begin{equation}}
\newcommand{\ee}{\end{equation}}
\begin{document}
 \title{(Nonequilibrium) dynamics of diffusion processes with non-conservative drifts.}
 \author{Piotr Garbaczewski  and Mariusz \.{Z}aba}
 \affiliation{Institute of Physics, University of Opole, 45-052 Opole, Poland}
 \date{\today }
 \begin{abstract}
   The nonequilibrium Fokker-Planck dynamics  with  a   non-conservative drift  field,  in  dimension $N\geq 2$,  {\it can be related  with the  non-Hermitian quantum mechanics in a real scalar potential $V$ and in a purely imaginary vector potential -$iA$ of real amplitude  $A$}, \cite{monthus}.
  Since  Fokker-Planck  probability density functions  may   be  obtained by means of Feynman's path  integrals,  the previous  observation points  towards a general issue of  "magnetically affine"    propagators, possibly    of quantum origin,  in real  and Euclidean  time.
  In below we shall follow   the  $N=3$  "magnetic thread", within which one may keep under a computational control   formally  and conceptually  different implementations of    magnetism (or surrogate magnetism)  in  the dynamics of  diffusion processes.   We shall focus on interrelations (with due precaution to varied, not evidently  compatible,  notational conventions) of:  (i) the pertinent non-conservatively drifted  diffusions, (ii) the classic  Brownian motion of charged particles  in the  (electro)magnetic field,
   (iii)  diffusion processes  arising  within   so-called Euclidean quantum mechanics (which from the outset employs  non-Hermitian "magnetic" Hamiltonians), (iv) limitations  of the usefulness of the  Euclidean map    $\exp(-itH_{quant}) \rightarrow \exp(-tH_{Eucl})$, regarding the probabilistic significance  of inferred (path) integral kernels in the description of diffusion processes.
  \end{abstract}
 \maketitle

\section{Introduction.}

\subsection{Preliminaries}

We are inspired by  the observation,  \cite{monthus}, that   the   dynamics of nonequilibrium diffusion processes in dimension $N\geq 2$  shows an affinity with   the "non-Hermitian electromagnetic quantum mechanics",  actually  appearing in its {\it fully}   Euclidean version, c.f. \cite{zambrini,zambrini1,gar,hasegawa} and \cite{avron,roep}.

 To set the ground for further discussion, we bring  to the reader's  attention  the origin of  the  transformation $-iA\rightarrow A$,  employed in \cite{monthus},  on the Lagrangian level of description of the   appropriate classical system  in real and Euclidean times respectively.

Namely, let us consider the classical Lagrangian  with the vector field entry (all dimensional constants have been  scaled away, to facilitate computations): ${\cal{L}}_{cl}(\vec{q}(t ), \dot{\vec{q}}(t)) =
{\frac{1}2}\dot{\vec{q}}^2(t)  + \dot{\vec{q}}(t)\cdot \vec{F}(\vec{q}(t))-  V(\vec{q}(t))$.  The Wick rotation $\tau =  it, t\geq 0$ leads us  to the complex-valued version of the  Lagrangian (remember about the need for an  overall sign change of the resultant expression) ${\cal{L}}_{Wick}(\vec{x}(\tau ), \dot{\vec{x}}(\tau )) =  - {\cal{L}}_{cl}(\vec{q}(t ), \dot{\vec{q}}(t)) =
{\frac{1}2}\dot{\vec{x}}^2(\tau )  - i \dot{\vec{x}}(\tau )\cdot \vec{F}(\vec{x}( \tau )) + V(\vec{x}(\tau ))$, with a purely imaginary vector entry $-i\dot{\vec{x}}\cdot \vec{F}(\vec{x})$.  The  subsequent  transformation $\vec{F} \rightarrow  -i \vec{F}$   gives rise to the real-valued version of the fully  Euclidean  Lagrangian  ${\cal{L}}_{Eucl}(\vec{x}(\tau ), \dot{\vec{x}}(\tau )) = {\frac{1}2}\dot{\vec{x}}^2(\tau )  -  \dot{\vec{x}}(\tau )\cdot \vec{F}(\vec{x}( \tau )) + V(\vec{x}(\tau ))$.  We indicate that no assumptions were made about  the  properties (conservative or not) of the vector field $\vec{F}$, which may as well be a pure gradient (e.g. conservative) one.

 We realise  that   for  "true" (like e.g. solenoidal)  magnetic potentials   $\vec{F} \equiv \vec{A}$ this would set a link with the Euclidean variant  of the (classical)  Maxwell theory, \cite{zampino,brill,zambrini}.  For future purposes we indicate that the transformation $\vec{A} \rightarrow -i \vec{A}$ replaces the $L^2$-Hermitian operator
$H_{quant} =-{\frac{1}2} (\vec{\nabla } - i\vec{A})^2$, \cite{grosche,glimm},  by the non-Hermitian one  $H_{Eucl} =-{\frac{1}2} (\vec{\nabla } - \vec{A})^2$, \cite{monthus}. On the level of    motion operators  (with scaled away physical constants), we ultimately arrive at the fully Euclidean  (but non-Hermitian)  outcome: $\exp(-itH_{quant})\rightarrow \exp(-\tau H_{quant}) \rightarrow \exp(-\tau H_{Eucl})$.

In passing we note  an  important sign difference, which needs to be accounted for, once we make a comparison with the reasoning of Ref.  \cite{zambrini}, where $H_{Eucl}$ and its $L^2$-adjoint $H_{Eucl}^*$ play a decisive role.  Namely, given the above  Hermitian generator $H_{quant}$,  a transformation  $\vec{A} \rightarrow +i \vec{A}$ employed in \cite{zambrini},   actually  produces   an operator  $H^*_{Eucl} =-{\frac{1}2} (\vec{\nabla } + \vec{A})^2$, which is the adjoint of  our  $H_{Eucl}$. This sign difference, while in a comparative vein, can be easily "corrected" by changing the sign of the vector potential in all relevant formulas of \cite{zambrini}.

\subsection{Drifted diffusion process.}

Our further attention will focus on    Markovian  diffusion processes driven by    non-conservative (generically non-gradient)   time-independent  drift fields $\vec{F}(\vec{x})$.  Let us consider a  diffusion process   $\vec{X}(t)$, associated with the stochastic differential equation of the Langevin-type, (interpreted in terms of infinitesimal  time increments)
 \be
d\vec{X}(t) = \vec{F}(\vec{X}(t)) dt + \sqrt{2\nu } d\vec{W}(t),
\ee
where a vector  field $\vec{F}(\vec{x})$  stands for a forward drift  of the process,   $\nu$  is  a  diffusion constant ($2\nu $ is  interpreted as the variance parameter), and  $\vec{W}(t)$  is the normalised   Wiener noise   in $R^3$, defined by   expectation values $\left< W_i\right> =0$ and $\left< W_j(s)W_j(t)\right> = \delta_{ij} \delta(s-t)$ for all $i, j =1,2,...N$.

Accordingly, if an initial probability density   function  $\rho_0(\vec{x})$ is given,   its  time evolution
$\rho_0(\vec{x})= \rho (\vec{x},0) \rightarrow  \rho (\vec{x},t)=  [\exp(tL^*)\rho_0](x)$  follows  the  Fokker-Planck equation \cite{wiegel,zaba,pavl}:
\be
\partial _t \rho = \nu \Delta \rho - \vec{\nabla }\cdot (\vec{F} \rho ) = L^* \rho ,
\ee
where the Fokker-Planck  operator $L^*=  \nu \Delta  - \vec{\nabla }\cdot (\vec{F} \cdot )$ is a  Hermitian $L^2(R^N)$  adjoint  of the  diffusion generator $L= \nu \Delta + \vec{F}\cdot \vec{\nabla}$, \cite{zaba,pavl}.  \\

  We  anticipate the existence of a transition probability density function  $p(\vec{y},s, \vec{x},t)$, $0\leq s<t\leq  T$, ($T\rightarrow \infty $ is admissible)  for the diffusion process (1), (2):   $\rho (\vec{x},t) = \int p(\vec{y},s, \vec{x},t) \rho (\vec{y},s) d^3y$, \cite{pavl,olk}.  We presume $p(\vec{y},s, \vec{x},t)$ to be a (possibly fundamental) solution of the F-P equation, with respect to  variables $\vec{x}$ and $t$, i.e.
\be
 \partial_t  p(\vec{y},s, \vec{x},t)  = L^*_{\vec{x}}p(\vec{y},s, \vec{x},t).
 \ee
   We  interpret   the transition pdf  as a (path)  integral kernel of the evolution operator  $\exp[(t-s)L^*]$, c.f. \cite{monthus,wiegel,hunt}, see also \cite{olk}. (Note that in the literature there exist other notational conventions for this pdf.)

In connection  with the  notion of the diffusion generator $L$, we indicate that  given  any  continuous and bounded function $f(\vec{x})$, we can introduce a function (actually a conditional expectation value of $f$, interpreted as an observable, \cite{pavl})
\be
u(\vec{y},s)  =  \mathbb{E}[f(\vec{X}_t)|\vec{X}_s =\vec{y}]=    \int f(\vec{x})  p(\vec{y},s, \vec{x},t) d^Nx
\ee
which solves a final value $u(\vec{y},s\rightarrow t)= f(\vec{y})$   parabolic problem  in  any  a priori  prescribed time interval $ s\in [0,t]$
\be
  -\partial_s u(\vec{y},s) = L_{\vec{y}} u(\vec{y},s),
\ee
with the  diffusion generator of  the form  $L =\nu  \Delta + \vec{F}\cdot \vec{\nabla}$.
Accordingly, \cite{pavl}, we have:
\be
- \partial_s  p(\vec{y},s, \vec{x},t) =L_{\vec{y}} p(\vec{y},s, \vec{x},t)
\ee
as a parabolic  companion of  Eq. (3) (albeit running in the reverse sense of time).
 Notice that equations (3) and (6) involve  non-Hermitian operators $L^*$ and $L$,  which are $L^2(R^3)$ adjoint to each other. This is a standard feature of Markovian diffusions.\\

{\bf Remark 1:} If the Fokker-Plack equation (2) admits an invariant (stationary) probability density   $\rho _*(\vec{x})$ as a solution, then the operators $L$ and $L^*$, which are non-Hermitian in $L^2$, become  Hermitian in function spaces $L^2_{\rho_*}$ and $L^2_{\rho_*^{-1}}$ respectively. Here,  $L^2_{\rho_*}$ indicates the Hilbert space , whose scalar product is weighted by $\rho _*$, according to $(f,g)_{\rho _*} = \int f(\vec{x})g(\vec{x}) \rho _*(\vec{x}) d^Nx$, and analogously for $L^2_{\rho_*^{-1}}$.

\subsection{"Magnetic" affinities.}

To conform with the notation of \cite{monthus,zambrini,gar,wiegel,hunt},  let us set $\nu =1/2$  (this amounts to rescaling the time label in the Fokker-Planck equation, cf. \cite{gar}).  By employing  the  identity  $\vec{\nabla }\cdot (\vec{F} \rho ) = (\vec{F}\cdot \vec{\nabla })\rho +
\rho (\vec{\nabla }\cdot  \vec{F})$  we arrive at   the following  form of the F-P operator $L^*$:
\be
L^* ={\frac{1}2} \Delta - \vec{F}\cdot \vec{\nabla} -  (\vec{\nabla }\cdot \vec{F}) =
{\frac{1}2}(\vec{\nabla } - \vec{F})^2  - {\cal{V}}  =  - (H_{Eucl} +{\cal{V}}) ,
\ee
where $H_{Eucl} =-{\frac{1}2} (\vec{\nabla } - \vec{F})^2$   and  $\cal{V}$ has a specific functional form
\be
 {\cal{V}} = {\frac{1}2} [(\vec{\nabla }\cdot\vec{F})+ \vec{F}^2],
\ee
which is  omnipresent in the discussion of diffusion processes in the  classical and quantum realm, \cite{zaba,olk,nelson}. For clarity of discussion, we reproduce
 an intermediate form  taken by  $L^*$ in the derivation of (7): $ L^* = [{\frac{1}2} \Delta - \vec{F}\cdot \vec{\nabla} -  (1/2)(\vec{\nabla }\cdot \vec{F})] + (1/2) \vec{F}^2] - {\cal{V}}$.

The diffusion generator $L$  reads
\be
L ={\frac{1}2} \Delta + \vec{F}\cdot \vec{\nabla} =  {\frac{1}2}(\vec{\nabla } + \vec{F})^2  -  {\cal{V}}  =  -(H^*_{Eucl} + {\cal{V}}),
\ee
with $H^*_{Eucl} =-{\frac{1}2} (\vec{\nabla } + \vec{F})^2$  and ${\cal{V}}$ given by  Eq. (8).    The intermediate form taken by $L$ reads: $ L = [{\frac{1}2} \Delta + \vec{F}\cdot \vec{\nabla} + (1/2)(\vec{\nabla }\cdot \vec{F}) + (1/2) \vec{F}^2]  - {\cal{V}}$.

We indicate  that for a divergenceless drift,  $div \vec{F} = \vec{\nabla }\cdot  \vec{F}= 0$, the F-P operator simplifies to   $L^*=(1/2) \Delta - \vec{F}\cdot \vec{\nabla}$, whose form is   fully   congruent with  that for  $L=(1/2) \Delta + \vec{F}\cdot \vec{\nabla}$. Consequently,  by merely  changing the sign of $\vec{F}$, we can map $L^*$ into $L$ and back.\\

The   above rewriting     of $L^*$ and $L$ is highly suggestive, \cite{monthus,zambrini,gar}, since non-Hermitian operators  $ H_{Eucl}= - (1/2)(\vec{\nabla } -  \vec{F})^2 $  and   $ H^*_{Eucl} = - (1/2)(\vec{\nabla } + \vec{F})^2$,  show a  Euclidean   affinity (normally restricted to magnetic potentials $\vec{F}\equiv \vec{A}$, \cite{monthus,zambrini})  with  (Hermitian)  quantum mechanical   magnetic Schr\"{o}dinger  operators, \cite{roep,avron}.

\subsection{Goals.}

We shall be interested in  interpreting the  transition probability densities of the diffusion processes in question, as  integral kernels  of the   motion operator   $\exp(tL^*)$:
\be
 p(\vec{y},s,\vec{x},t)= [e^{L^*(t-s)}](\vec{y},\vec{x}) = [e^{-(H_{Eucl} + {\cal{V}})(t-s)} ](\vec{y},\vec{x}).
\ee

It is a classic observation,  \cite{wiegel,hunt}, that  Fokker-Planck  transition probablity density functions  and probability densities,  for diffusions with non-conservative drifts, are    amenable to   Feynman's  path integration routines. In case of  conservative drifts, the same goal can be achieved  by means of a   multiplicative (Doob-like)  conditioning of the related (positive)   Feynman-Kac kernel,  \cite{gar,roep,olk,gar3,zaba,glimm}, provided the existence of   stationary  pdfs is granted.

  In the  path integral derivation,  we need to define the action functional in terms of a suitable  Lagrangian, with the obvious advantage  that in the  quadratic  case the classical action alone  normally  suffices  for the evaluation of the path integral, c.f. \cite{roep,grosche,glimm}, see also  \cite{glasser,pell}. By this reason, a substantial part of the  discussion in  Ref. \cite{monthus},   has been  dedicated   to  linear   drift fields, with   emphasis on the existence of nonequilibrium  steady states $\rho_*(\vec{x})$, and   non-vanishing steady  currents $\vec{j}_*(\vec{x},t)$.

 Our major aim is to  examine   the  $N=3$  "(electro)magnetic thread", which  comprises  conceptually  different  "(electro)magnetic"  implementations of drifted diffusion processes. This amounts to revisiting:
   (i) the  classic   theory of  the Brownian motion in a magnetic field, \cite{czopnik,czopnik1} and \cite{aquino,aquino1,abdoli}, (ii)  past research on    path integral solutions of the Fokker-Planck equation for a system with non-conservative forces, \cite{wiegel,hunt}, (iii)the  Euclidean time dynamics, generated by Hermitian Schr\"{o}dinger operators in magnetic fields and path integral kernels of related    Schr\"{o}dinger semigroups,  \cite{avron,roep,glimm},  (iv)  electromagnetic  dynamical features deduced within    so-called Euclidean quantum mechanics, \cite{zambrini,zambrini1} (see also \cite{hasegawa,morato,nelson} for a complementary  analysis of the  Lagrangian  variational  principle with a classical action).

  In particular, we aim at a  clear  discrimination  between the physically motivated  impact of magnetism on diffusing charges,   and "electromagnetic" analogies   appearing in the study of nonequilibrium  diffusion processes  with non-conservative drifts.  The latter, may    not necessarily  embody  the very concept of   diffusing  charged particles, but  nonetheless  might satisfactorily  mimic (simulate) the electromagnetic-looking   behavior of diffusion currents,  in terms of  a potentially useful  "surrogate  magnetism".

 We shall derive a number of exact transition pdfs  for the pertinent  spatial  diffusion processes. While departing from    phase-space   derivations  of Refs. \cite{czopnik,czopnik1}, we may  extract    formulas  which   can be   reinterpreted solely  in the  configuration space  (via fine tuning and scaling away  of various parameters,  and temporarily suspending the involved fluctuation-dissipation relationship).

In the course of our discussion,  we perform a detailed derivation of   the   propagator associated with     would-be  natural Euclidean analogue $H_{Eucl} = -(1/2)(\vec{\nabla } -\vec{F})^2$   of   $H_{quant} = - (1/2) (\vec{\nabla }+  i\vec{F})^2$, for a solenoidal drift field $\vec{\nabla} \cdot \vec{F}=0$. This is accomplished   by a  direct  evaluation of the   path  integral  kernel of $\exp(-tH_{Eucl})$.

Our motivation comes from  the fact that the   integral kernel of the  legitimate  (Hermitian) Schr\"{o}dinger semigroup  $\exp(-t H_{quant})$  is complex-valued.  We have verified  that the kernel of $\exp(-t H_{Eucl})$ is real-valued, but shows a number of deficits (detected before in Ref. \cite{gar}) which preclude its unrestricted   usefulness  {\it as it stands}  for the  description of diffusion processes.

 In passing, we mention that the  complex-valued  integral kernel of $\exp [-\beta H_{quant}]$, with $\beta \sim 1/kT$,  has an  interpretation  as the unnormalized density matrix (which in turn yields a  legitimate  partition function of the   quantum statistical system)  in the classic study of  the diamagnetism of free electrons, \cite{sond,glasser}.

\section{Signatures of nonequilibrium.}

Concerning the nonequilibrium properties of diffusion processes in $N\geq 2$, we   restrict our attention to  relaxation  scenarios,  within which  $\rho(\vec{x},t)$  asymptotically  (as $t\to \infty $) approaches a  stationary  (steady state) probability  density $\rho _*(\vec{x})$. If  $D\geq 2$ such pdf  may  in principle   coexist with  the  non-vanishing steady  current  $j_*(\vec{x}) \neq 0$, c.f. \cite{monthus}.  To this end we assume  form the start,  that  drift fields $F(\vec{x})$ are  non-conservative, i.e.  cannot be represented in  the pure  gradient form.

The  diffusion current notion appears through  rewriting  the Fokker -Planck equation $\partial_t \rho = (1/2)\Delta \rho - \vec{\nabla} \cdot (\vec{F}\cdot  \rho )$  as   the continuity equation  for $\rho (\vec{x},t)$:
\be
\begin{split}
\partial _t \rho = - \vec{\nabla }\cdot  \vec{j} =  - \vec{\nabla }\cdot ( \vec{v}  \rho),\\
 \vec{v}  =   \vec{F} - \vec{\nabla }\ln \rho ^{1/2},\\
 \end{split}
\ee
where      $\vec{v}$ is  a current velocity field.

Let us assume that the Fokker-Planck equation  (1)  admits   a stationary  pdf $\rho_*(\vec{x})$. In view of $\partial _t\rho _*=0$, the   related  asymptotic   diffusion current $\vec{j}_*=   \rho _*  \vec{v}_* $, with
$\vec{v_*}  =   \vec{F} - \vec{\nabla }\ln \rho_* ^{1/2}$   needs either to vanish, $\vec{j}_*(\vec{x})=0$, or to be  divergenceless,  $\vec{\nabla }\cdot  \vec{j}_* =0$.

 The    choice of the drift field in   gradient  form $\vec{F} =  \vec{\nabla }\ln \rho _* ^{1/2}$, compare e.g.  Ref.\cite{zaba},   would  secure a relaxation property $\rho (\vec{x},t) \rightarrow \rho _*(\vec{x})$, with no steady  current at all, since  then  $\vec{j}_*=0$  identically.

 By denoting  $\rho _* = \exp(-2 \phi )$, (that amounts to $\rho _*^{1/2}= \exp(-\phi)$, the notation predominantly used in \cite{pavl,zaba})   we are left with the identity
 $\partial _t\rho _*= -\vec{\nabla }\cdot [\rho _*  (\vec{F}  +   \vec{\nabla }\phi )] = 0$.  The  non-vanishing steady current  $ \vec{j}_*\neq 0$  may   coexist with $\rho _*$,  if   the drift field  $\vec{F}$  is selected as the non-gradient   one.

 Let us consider  non-conservative drifts, which  decompose  into a sum  comprising any (no necessarily related to $\rho _*$)  gradient entry  $ - \vec{\nabla }\phi $ ,   and the non-gradient one  $\vec{A}$   (this notation sets  a  correspondence with  a magnetic  vector  potential in  $R^3$). We have
   \be
\begin{split}
   \vec{F}= \vec{A} -  \vec{\nabla }\phi  \Longrightarrow \\
 {\cal{V}}= {\frac{1}2} [(\vec{\nabla \phi  })^2- \Delta \phi ] + {\frac{1}2} [\vec{A}^2 + {\vec{\nabla }}\cdot \vec{A}]  - \vec{A} \cdot {\nabla }\phi ,\\
\end{split}
 \ee

Assuming    $-\phi =  \ln \rho _*^{1/2}$, we realise that
 the   steady   diffusion current   $\vec{j} = \vec{A} \rho _*$  must  be divergenceless. Accordingly,
 \be
0= \vec{\nabla }\cdot(\vec{A} \rho _*)=  (\vec{A}\cdot \vec{\nabla })\rho _*  +  \rho_* (\vec{\nabla }\cdot  \vec{A})\Longrightarrow
\vec{\nabla }\cdot  \vec{A} =  2 \vec{A}\cdot \vec{\nabla } \phi .
\ee
In view of (13),  the magnetic contribution to ${\cal{V}}$  in (12)  reduces to ${\frac{1}2} \vec{A}^2$.
If we  additionally  assume   that $\vec{A}$ is a solenoidal vector field,   $\vec{\nabla }\cdot \vec{A}=0 $ ,
we arrive at the orthogonality relation $\vec{A}\cdot \vec{\nabla } \phi =0$,  valid  for all $\vec{x} \in R^N$.
This correlates $\vec{A}$ with $\vec{\nabla }\phi $.

In connection with (12),  we realize that if  we skip $\phi $ (set $\phi \equiv 0$  or equal to any constant), we stay with $\vec{F}= \vec{A}$   and ${\cal{V}} = {\frac{1}2}[\vec{A}^2 + \vec{\nabla}\cdot \vec{A}]$, which reduces to $ {\frac{1}2}\vec{A}^2$ for solenoidal   and constant  vector fields.  On the other hand, if we disregard $\vec{A}$ by setting $\vec{A}\equiv \vec{0}$, we are left with $\vec{F}= - \vec{\nabla }\phi $  and ${\cal{V}}= {\frac{1}2} [(\vec{\nabla \phi })^2- \Delta \phi ] $.\\

{\bf Remark 2:} The    "magnetic affinity"  (7)-(9) acquires  a deeper meaning,  if  the  drift field   $\vec{A}(\vec{x})$  induces   the  non-vanishing  "magnetic matrix" $\textbf{B}[\vec{A}]: B_{ij} = \partial_iA_j - \partial_jA_i$, with    $1\leq i,j \leq N$,   $\textbf{B}\neq \textbf{0}$. This implies
$\textbf{B}[\vec{F}]  \neq \textbf{0}$ as well.   We note that for well behaved functions  $\phi (\vec{x})$  (with  continuous  mixed derivatives),  $\textbf{B}[\vec{\nabla} \phi ]= \textbf{0}$. Accordingly, without the $\vec{A}$ contribution  in Eq. (12) we are left with a  conservative vector field  $\vec{F}= - \vec{\nabla}\phi $.
Since  conservatively drifted diffusions have received an ample coverage in the literature, c.f.  \cite{pavl} and \cite{zaba}, we shall focus our attention on the non-conservative ones.\\

In $N=3$ dimensions, the non-conservativeness of the vector field $\vec{F}$  stems from its non-vanishing curl,   $curl \vec{F} =\vec{\nabla } \times \vec{F}\neq \vec{0}$. For the  conservative (i.e.  gradient) drift  $\vec{F} \sim  - \vec{\nabla } \phi $  we have $\vec{\nabla }\times (\vec{\nabla }\phi )=0$.

The non-conservative field  itself $\vec{F}$   needs not to be purely rotational and   may encompass both the gradient and  rotational contributions on an equal footing, like in   a decomposition  (12),  c.f. \cite{monthus,zambrini,gar}), $ \vec{F} = \vec{A} - \vec{\nabla } \phi $, where $\vec{A}$ is considered to be purely rotational.
For  well behaved  $\phi$  we have  $ \vec{\nabla }\times  (\vec{F} - \vec{A}) = 0$, in affinity with the standard   gauge transformation of  any   electromagnetic potential, which  leaves the Lorentz force intact, c.f.  \cite{roep} and  check  the Lorentz force appearance in the  Brownian motion, \cite{czopnik,czopnik1,aquino,aquino1,abdoli}.

Would we have left $\phi $ aside by setting $\phi =0$, the resultant drift field would reduce to   $\vec{F} = \vec{A}$, with the path-wise dynamics (1), driven by the purely rotational (like solenoidal)  vector field, c.f \cite{hasegawa,gar}.

\section{Path integration hints.}

\subsection{Lagrangian dynamics may look "electromagnetic".}

A distinctive research topic   in Ref. \cite{monthus} has been  a  discussion of  general  path   integral formulas  for  propagators (fundamental solutions, transition probability densities) of  Fokker-Planck equations  with  non-conservative  drifts. That  basically   refers to    quadratic Lagrangians, with semiclassical options for a direct computability of involved path integrals, which incorporate $N\geq 3$ analogues of the  familiar $N=3$ magnetic coupling. See e.g. also \cite{wiegel,hunt}, and  classic derivations of path integral formulas for  propagators of    "magnetic" Schr\"{o}dinger equations, analytically continued to  "imaginary time", \cite{avron,roep,gar,glasser}.

The   path integral context  for non-conservatively drifted diffusion processes   has been  set in Refs. \cite{wiegel,hunt}, and revived in \cite{monthus},   through the formula   "for the propagator associated with the Langevin system" (actually  the  integral kernel  $\exp(tL^*)(y,x)$ of the operator $\exp(tL^*)$:
\be
p(\vec{y},0,\vec{x},t)=  \exp(L^*t)(\vec{y},\vec{x})=  \int_{\vec{x}(\tau =0)=\vec{y}}^{\vec{x}(\tau =t)=\vec{x}}
   {\cal{D}}\vec{x}(\tau ) \,  \exp \left[ - \int_0^t  d\tau {\cal{L}}(\vec{x}(\tau ), \dot{\vec{x}}(\tau )) \right],
\ee
where the $\tau $-dynamics stems  from  the Euclidean  Lagrangian ${\cal{L}}_{Eucl}$ (cf.  Section I.A and  \cite{monthus,wiegel,hunt}),  hereby denoted ${\cal{L}}$:
\be
{\cal{L}}(\vec{x}(\tau ), \dot{\vec{x}}(\tau )) = {\frac{1}2} \left[ \dot{\vec{x}}(\tau ) - \vec{F}(\vec{x}(\tau ))\right]^2  + {\frac{1}2}  \vec{\nabla }\cdot \vec{F}(\vec{x}(\tau ))=
{\frac{1}2}\dot{\vec{x}}^2(\tau )  - \dot{\vec{x}}(\tau )\cdot \vec{F}(\vec{x}(\tau )) + {\cal{V}}(\vec{x}(\tau )),
\ee
with ${\cal{V}}(\vec{x})$ given by  Eq. (8).

 We recall  that the "normal"   (e.g. non-Euclidean)  classical Lagrangian would have the form $L = T - V$ with $T=  \dot{\vec{x}}^2 /2$   and $V(\dot{\vec{x}},\vec{x},t)= {\cal{V}}   - \dot{\vec{x}}\cdot \vec{F}$. See e.g. the  Appendix A in Ref. \cite{monthus} and \cite{nelson}. Note that the diffusion-induced  Lagrangian (15) actually has the  Euclidean  form   ${\cal{L}}= {\cal{T}} + V$. This  has consequences for    the derived versions of the second Newton law (e.g. the sign of the derived Lorentz force  analogue, in this connection see Remark  in below).\\

Since we have  in  hands  an explicit Lagrangian (15), (while keeping  in memory its relevance for the evaluation of  path integrals in the  quadratic case), we may ask for the  dynamical output in terms of the Euler-Lagrange equations, still without specifying detailed properties of the   vector field $\vec{F}(\vec{x}(t), t)$, except for tentatively  admitting a direct dependence  on time.  To compress the resulting formulas we pass to the notation $\vec{x}= (x_1,x_2,x_3)$, so that   $V(x, \dot{x},t) = {\cal{V}}(x,t)- \sum_j \dot{x}_j F_j(x,t)$, c.f. \cite{nelson}.   The  Euler-Lagrange equations for   the Lagrangian  ${\cal{L}} = {\cal{T}} + {\cal{V}}$,   read:
\be
{\frac{\partial {\cal{L}}}{\partial x_i}} - {\frac{d}{dt}} {\frac{\partial {\cal{L}}}{\partial \dot{x}_i}}=0 \Longrightarrow
{\frac{\partial V}{\partial x_i}}  - {\frac{d}{dt}} \left( {\frac{\partial {\cal{T}}}{\partial {\dot{x}}_i}}  +  {\frac{\partial V}{\partial {\dot{x}}_i}}\right) = 0
\ee
for  $i =1,2,3$. Accordingly, we have
\be
\ddot{x}_i = \left({\frac{\partial {\cal{V}}}{\partial {x_i}}}  + {\frac{\partial F_i}{\partial t}}\right) + \sum_j  B_{ij} \dot{x}_j,
\ee
where the notation $B= (B_{ij})$   refers to the antisymmetric matrix:
\be
B_{ij}= {\frac{\partial F_i}{\partial x_j}} -  {\frac{\partial F_j}{\partial x_i}}
\ee
named in Ref. \cite{monthus} a magnetic matrix, c.f.  our discussion following Eq. (13).

In $N=3$ dimensions, nonvanishing   components of the magnetic matrix, actually  define the magnetic  vector
\be
\vec{\nabla } \times \vec{F}  =\vec{B}= (B_1=B_{32},B_2= B_{13},B_3=B_{21})= (\partial_2F_3 -\partial_3F_2,\partial_3F_1 - \partial_1F_3,\partial_1F_2-\partial_2F_1).
\ee
Denoting   $\sum_j  B_{ij} \dot{x}_j= F^{magn}_i$, we  realize that
\be
\vec{F}^{magn}=- \dot{\vec{x}}\times (\vec{\nabla } \times \vec{F})= - \dot{\vec{x}} \times \vec{B},
\ee
as required (up to a sign, which is opposite to that in the "classical" case) from the magnetic part of the Lorentz force.

 The electric analogue of this force reads, c.f. Eq. (16)  $\vec{F}^{el} = \vec{\nabla} {\cal{V}}  +  {\partial \vec{F}}/\partial t$   and is opposite to that valid  in the "classical" case.

  We note that the  decomposition   of the derived   Euclidean Lorentz force  $\vec{F}^{Lorentz}  = \vec{F}^{el}+\vec{F}^{magn}$ into a sum of electric and magnetic contributions is not "clean". The "electric" term has explicit  $\vec{A}$-dependent  contributions, through $- \vec{\nabla } {\cal{V}}$ and  ${\partial \vec{F}}/\partial t$.

In passing we note that the derived "magnetic" force is  always  orthogonal to the velocity, since $\vec{F}^{magn}\cdot \dot{\vec{x}} =\sum_{ij} B_{ij} \dot{\vec{x}}_i \dot{\vec{x}}_j =0$.\\

  To have a clear  view  of the above  Euclidean-looking  analogue  of the standard Lorentz force, let us invoke the text-book wisdom, \cite{cohen}.  Namely,  the standard ("classical") Lorentz force expression  for  a particle of mass $m$ and charge $q_c$ (any sign and size):
\be
 \ddot{\vec{x}} = {\frac{q_c}m}  [\vec{E}(\vec{x},t)  + \dot{\vec{x}}\times \vec{B}(\vec{x},t)]
\ee
derives from the classical  Lagrangian (where we  temporarily  restore relevant dimensional constants)
\be
{\cal{L}}_{cl}(\vec{x}, \dot{\vec{x}},t) = {\frac{1}2}
m\dot{\vec{x}}^2 + q_c \dot{\vec{x}}\cdot \vec{A}(\vec{x},t)  - q_c U(\vec{x},t).
\ee
We note that by  passing from ${\cal{L}}_{cl}$ to $(1/m){\cal{L}}_{cl}$   and absorbing the $q_c/m$ coefficient in the redefined  functions
 $(q/m) \vec{A}= \vec{F}$ and $(q/m)U= {\cal{V}}$, we  recover
\be
{\cal{L}}_{cl}= {\frac{1}2}\dot{\vec{x}}^2  - [{\cal{V}} - \dot{\vec{x}}\cdot \vec{F}] = T- V  \Longrightarrow
-  {\frac{\partial V}{\partial x_i}}  - {\frac{d}{dt}} \left( {\frac{\partial {\cal{T}}}{\partial {\dot{x}}_i}}  -  {\frac{\partial V}{\partial {\dot{x}}_i}}\right) = 0
\ee
with  $V={\cal{V}} - \dot{\vec{x}}\cdot \vec{F}$. This implies the   sign inversion of the inferred  "classical"  Lorentz force expression
\be
\ddot{x}_i =    - \left[ \left({\frac{\partial {\cal{V}}}{\partial {x_i}}}  + {\frac{\partial F_i}{\partial t}}\right) + \sum_j  B_{ij} \dot{x}_j  \right],
\ee
if compared with the  "nonclassical"  outcome (17)  (devoid of dimensional constants).

 Accordingly, the two formalisms lead to  the  opposite (in sign) Lorentz forces, if evaluated for the very same charge.  That in view of the sign difference  on the right-hand side of the  Euler-Lagrange equations (17) and (23).\\

 {\bf Remark 3:} We point out the  gauge invariance of both the electric and magnetic fields, and thence the Lorentz force expressions, while to the contrary,  the stochastic differential  equation (1), and generators   (7),  (9), are sensitive to the choice of gauge.   Namely, once we have  given the drift field $F(\vec{x},t)$,  and  an (electric)   potential $\phi (\vec{x},t)$, we can pass to  the gauge transformed pair of functions   $\vec{G}  =  \vec{F} + \vec{\nabla } \eta (\vec{x},t)$   and  $\Phi = \phi -  \partial _t \eta (\vec{x},t)$, where $\eta  (\vec{x},t)$ is any scalar function. These  functions    determine the  very same  $\vec{E}= - \vec{\nabla} \Phi  - \partial _t \vec{G}$ and $\vec{B}= \vec{\nabla } \times  \vec{G}$ as   the   former  pair    $\vec{F}$  and $\phi $, \cite{cohen}.

\subsection{Lagrangian signatures of stationary pdfs.}

Let us consider the action functional  (e.g. minus exponent) in Eq. (14),  in association with the  drift field (12), where $- \phi = \ln \rho _*^{1/2}$.   By  following arguments of  Section II.D of Ref. \cite{monthus} (while adopted to our notation), we readily infer that the term $\dot{\vec{x}}\cdot \vec{F}$ in the Lagrangian (15) contributes:
\be
\int_0^t \dot{\vec{x}}\cdot [- \vec{\nabla} \phi (\vec{x}(\tau ))  + \vec{A}(\vec{x}(\tau ))]  d\tau =
  - \int_0^t {\frac{d}{d\tau }} \phi (\vec{x}(\tau )) d\tau  + \int_0^t \dot{\vec{x}}\cdot  \vec{A}(\vec{x}(\tau ) ) d\tau = \phi (\vec{x}(0)) - \phi (\vec{x}(t))  + \int_0^t \dot{\vec{x}}\cdot  \vec{A}(\vec{x}(\tau )) d\tau
\ee
to the action functional.

Thence, the related probability density function  (path integral kernel of $\exp(tL^*)$) should arise in the form:
\be
p(\vec{y},0,\vec{x},t)=  e^{\phi (\vec{y}) - \phi (\vec{x})}\,  k(\vec{y},0,\vec{x},t)
\ee
where the new function $k(\vec{y},0,\vec{x},t)$ is no longer a transition probability density (does not integrate to one) but an integral kernel (e.g. the propagator) of another  motion operator (to be identified in below):
 \be
 k(\vec{y},0,\vec{x},t) =   \int_{\vec{x}(\tau =0)=\vec{y}}^{\vec{x}(\tau =t)=\vec{x}}
   {\cal{D}}\vec{x}(\tau ) \,  \exp \left[ - \int_0^t  d\tau {\cal{L}}_{magn}(\vec{x}(\tau ), \dot{\vec{x}}(\tau )) \right].
\ee

By employing  the property (13) (the existence condition for steady current, c.f. Section II.D in \cite{monthus}),  we arrive at
\be
\begin{split}
 {\cal{V}} = V  + {\frac{1}2}\vec{A}^2,\\
 V(\vec{x})= {\frac{1}2} [(\vec{\nabla } \phi  )^2  - \Delta \phi ],\\
 \end{split}
\ee
where the functional  form of $V$ is a consequence of our gradient choice $F= - \nabla \phi $ for the drift field.

In the above,
\be
{\cal{L}}_{magn} (\vec{x}(\tau ), \dot{\vec{x}}(\tau )) =
{\frac{1}2}\dot{\vec{x}}^2(\tau ) - \dot{\vec{x}} \cdot \vec{A}(\vec{x}(\tau )) +  {\cal{V}}(\vec{x}(\tau )) =
{\frac{1}2} \left[ \dot{\vec{x}}(\tau ) - \vec{A}(\vec{x}(\tau ))\right]^2  + V(\vec{x}(\tau )).
\ee
On the level of operators, the passage from the  transition kernel $p$ of  (26) to $k$ of (27), amounts to the similarity transformation, discussed in Section II.E of Ref. \cite{monthus}, c.f. also for  analogous  considerations  (pertaining to conservative gradient drifts) in Section 2.4 of Ref. \cite{zaba}:
\be
H_{magn} = e^{\phi } L^* e^{-\phi }=  - {\frac{1}2} (\vec{\nabla } - \vec{A})^2 + {\cal{V}}.
\ee
The outcome can be readily verified by resorting to the operator identity  $e^{\phi } \vec{\nabla } e^{- \phi } =  \vec{\nabla } - (\vec{\nabla }\phi )$.

Returning back to the formula (26), we  readily  recognize the term    $\exp (-tH_{magn})(\vec{y},\vec{x}) = k(\vec{y},0,\vec{x},t)$, whose  path integral evaluation involves ${\cal{L}}_{magn}$, c.f (27) - (30).

Let us disregard $\vec{A}$ contributions and  consider $\vec{F}= - \vec{\nabla }\phi $   in the formulas (14) and (25).  We  readily arrive at the factorised  transition probability density of the   conservatively-drifted  diffusion process, whose form is analogous to (26). The integral kernel $k(\vec{y}, 0,\vec{x}, t)$ takes the  form
\be
k_{st}(\vec{y},0,\vec{x},t) =   \int_{\vec{x}(\tau =0)=\vec{y}}^{\vec{x}(\tau =t)=\vec{x}}
   {\cal{D}}\vec{x}(\tau ) \,  \exp \left[ - \int_0^t  d\tau {\cal{L}}_{st}(\vec{x}(\tau ), \dot{\vec{x}}(\tau )) \right],
\ee
where (c.f. Eq. (15) and note that $V$ replaces  ${\cal{V}}$.)
\be
\begin{split}
{\cal{L}}= {\cal{L}}_{st} + \dot{\vec{x}}\cdot \vec{\nabla}\phi ,\\
                {\cal{L}}_{st} (\vec{x}(\tau ), \dot{\vec{x}}(\tau )) =
{\frac{1}2}\dot{\vec{x}}^2(\tau ) +  V(\vec{x}(\tau )),\\
\end{split}
\ee
and  ${\cal{L}}_{st}$  is known to  give rise to the standard  Feynman-Kac propagator, see e.g \cite{glimm,roep,pavl,olk,gar3,zaba}.

  On the operator level, one encounters  the  similarity transformation of the form (30),  provided all $\vec{A}$ contributions are neglected.  One arrives at
   \be
   H_{st} = e^{\phi } H e^{-\phi }=  - {\frac{1}2} \Delta   +  V.
    \ee
  The pertinent transformation is discussed in minute detail in Ref. \cite{pavl}, see also \cite{zaba}. \\

  We indicate that in principle, we  can   consider a purely conservative  drift term  $\vec{F} = - \vec{\nabla} \phi $ in the  Fokker-Planck equation (1). Then, the  auxiliary potential ${\cal{V}}(\vec{x})$, Eq. (8),   takes the  form   (28) with no magnetic contribution.  The notation $V$ instead of ${\cal{V}}$  of Eq. (28),  distinguishes  the   gradient drift-induced   version of  the potential  from  the   general form (8).

    If  $\rho _* = \exp(-2 \phi )$, we have  $\vec{F}= \vec{\nabla } \ln \rho _*^{1/2}= - \vec{\nabla \phi}$,  \cite{olk,gar3,zaba} and the above   potential $V$ can be given another   form, \cite{zaba,gar4},  $V= (1/2) \Delta \rho _*^{1/2}/ \rho _*^{1/2}$,   which notoriously  reappears in the in the literature on confined diffusion processes  in classical and quantum contexts,    \cite{zambrini,gar,zaba,hasegawa,morato,nelson,gar4}.\\

{\bf Remark 4:}  We note that the transition probability density (26), with the Feynman-Kac kernel entry (31), can be derived directly from (14), by inserting the Lagrangian ${\cal{L}}= {\cal{L}}_{st} + \dot{\vec{x}}\cdot \vec{\nabla}\phi $ instead of (15).
By formally choosing $\phi (\vec{x}) = \vec{x}^2/2$, we can rewrite ${\cal{L}}$ in the form  ${\cal{L}} = {\frac{1}2} (\dot{\vec{x}} + \vec{x})^2  -  {\frac{3}2}$.  Remembering  that  $\vec{F} = - \vec{\nabla }\phi = -\vec{x}$, we   hereby  restore    the primary form (15)  of the Lagrangian  ${\cal{L}} = {\frac{1}2} (\dot{\vec{x}} - {\frac{1}2}\vec{F})^2  +  \vec{\nabla }\cdot \vec{F}$, while adopted to our special (conservatively drifted)  case.
For  completeness, we note that  the Fokker-Planck operator  $L^*$  related to the pdf (26),  directly stems from (7)  and takes the form $L^*= {\frac{1}2} \Delta  + \vec{x}\cdot \vec{\nabla } + 3= - [-{\frac{1}2}[(\Delta  + \vec{x})^2 + V] $, with $V= (\vec{x}^2 - 3)/2$.

\section{A  detour: Brownian motion in a magnetic field.}
\subsection{Frictionless regime.}

 The original study of  Sect. 3 in  Ref. \cite{czopnik}  refers to a phase-space description of the  damped random motion of a  charged particle (with charge $q_e$) in a constant magnetic field and fluctuating electric environment. In contrast with  \cite{czopnik}, in Ref. \cite{czopnik1}  the frictional contribution has been skipped. The magnetic field has been presumed to be constant and oriented in the $z$-direction  of the Cartesian reference frame,   $\vec{B} =(0,0,B)$.

  The  frictionless  Langevin-type equation  in   velocity space has been considered as the  randomized version of the second Newton   law,  with the magnetic part of the Lorentz force  subject to  random  fluctuations, in conformity  with  standard white-noise statistics assumptions. For clarity of discussion, we reproduce the   version employed in Ref. \cite{czopnik}:
 \be
{\frac{d\vec{u}}{dt}} = {\frac{q_{c}}{mc}} \vec{u} \times \vec{B}   +   \vec{{\cal{E}}}(t).
 \ee
 Here ${\cal{E}}$ {\it is not } an electric part of the Lorentz form but represents the  white  noise term,  which we traditionally interpret  in terms of the statistics of the Wiener process increments, while rewriting Eq. (34) in the form analogous to (1) (up to the proper  adjustment of parameters)
\be
d\vec{u}(t) =  \vec{F}(\vec{u}(t))  dt + \sqrt{2q } d\vec{W}(t)
\ee
with a diffusion constant $q$ and    the forward  drift   given as  a vector field in  velocity space
\be
\vec{F}(\vec{u}(t)) = - \Lambda \vec{u}(t),
\ee
where
\be
\Lambda =
\left(
  \begin{array}{ccc}
    0 & - \omega _c & 0 \\
    \omega _c & 0 & 0 \\
    0 & 0 & 0 \\
  \end{array}
\right)
\ee
and the   cyclotron frequency parameter $\omega_c$ reads $\omega_c = q_cB/mc  $.

In Refs. \cite{czopnik,czopnik1} one can find detailed derivations, which we  skip in the present paper. An important outcome is that  any  appropriate probability density function  $\rho(\vec{u},t)$ is a solution of the Fokker-Planck equation of the form   (2):
\be
\partial _t \rho = q \Delta _{\vec{u}} \rho      - \omega _c \left[\nabla _{\vec{u}}\times {\vec{u} \rho }\right]_{i=3}  =q \Delta _{\vec{u}} \rho  - (\vec{\nabla }_{\vec{u}}\cdot \vec{F})  \rho  = L^*_{\vec{u}} \rho ,
\ee
where
\be
\vec{F}(\vec{u}) =  \omega _c (u_2,-u_1,0).
\ee
The underlying  Markovian process in the velocity space  is  time   homogeneous, and  has a transition probability density  obeying the initial condition  $p(\vec{v}_0,0, \vec{u},t) \rightarrow \delta ^3(\vec{u}- \vec{v}_0)$ as $t\rightarrow 0$:
\be
p(\vec{v_0},0,\vec{u},t)  = \left( {\frac{1}{4\pi qt}}\right)^{3/2}  \, \exp \left(  -  {\frac{(\vec{u} - U(t)\vec{v}_0)^2}{4qt}}\right).
\ee
Here, the rotation matrix   $U(t)= \exp ( - t\Lambda )$ reads:
\be
U(t) = \left(
         \begin{array}{ccc}
          cos(\omega _ct) &sin(\omega _ct)&0\\
           -sin(\omega_ct) & cos(\omega _ct) &0\\
           0&0& 1\\
         \end{array}
       \right).
\ee
The transition pdf (40) is a solution of the Fokker-Planck equation (38).  An extension of  $p(\vec{v_0},0,\vec{u},t)$ to the general form of the transition pdf  $p(\vec{v},s,\vec{u},t)$  is immediate. The  time-homogeneity of the  process  implies that we can replace the $t$ by  $(t-s)$  label, in conjunction with the replacement of  $\vec{v}_0$   by $\vec{v}$,   in the formula (40).

\subsection{Reintroducing friction.}

The frictionless diffusion described above, can be deduced from the general phase-space formalism,  set for the Brownian motion in a magnetic field \cite{czopnik}, as long as we do not insist on the validity of the fluctuation-dissipation relationships, and stay on the velocity space level of the analysis.

We begin from a redefinition of the Langevin-type equation (35) to encompass the simplest damping scenario
\be
{\frac{d\vec{u}}{dt}} =   - \beta  \vec{u}  + {\frac{q_{c}}{mc}} \vec{u} \times \vec{B}   +   \vec{{\cal{E}}}(t),
\ee
where $\beta >0$ stands for a friction coefficient.   In (42)  we  assume $\vec{B}=(0, 0, B)$,  hence  a rewriting  (36), (37)  is still valid, except for another form of the matrix $\Lambda$, which reads:
 \be
\Lambda =
\left(
  \begin{array}{ccc}
    \beta  & - \omega _c & 0 \\
    \omega _c & \beta  & 0 \\
    0 & 0 & \beta \\
  \end{array}
\right).
\ee
Accordingly (we note that $\Lambda $ can be decomposed into a sum o commuting matrices):
\be
e^{- t\Lambda}  = e^{- \beta t} U(t)
\ee
with $U(t)$ given by Eq. (41).

The corresponding  Fokker-Planck equation  retains the  functional form (35), but $\vec{F}$  presently looks otherwise
\be
\vec{F}(\vec{u}) = (\omega _c u_2  - \beta u_1, -\omega _c u_1 - \beta u_2,  - \beta u_3),
\ee
and the previous purely  rotational  drift   $\vec{\nabla} \cdot \vec{F}= 0$  has been modified to yield     $\vec{\nabla} \cdot \vec{F}= - 3\beta $.
The vector field $\vec{F}$,  Eq. (36), can be written as a sum of purely  rotational and conservative (while in velocity space) contributions, c.f. Eq. (12).\\

The transition probability  density of the  frictional diffusion process, conditioned by the
initial data $u_0$  at $t_0 =0$ reads:
\be
p(\vec{v_0},0,\vec{u},t)  = \left( 2\pi {\frac{q}{\beta}}(1- e^{-2\beta t})\right)^{- 3/2}  \,
 \exp \left( - {\frac{(\vec{u} - e^{-\beta t} U(t)\vec{v}_0)^2}{2 {\frac{q}{\beta }}(1- e^{-2\beta t})}}\right).
\ee
The process is time-homogeneous, hence the above formula in fact defines $p(\vec{v},s,\vec{u},t)= p(t-s,\vec{v},\vec{u}) $.

For any positive value of $\beta$, an asymptotic stationary pdf (stationary solution of the Fokker-Planck equation)
has the form (we follow the notation of Ref. \cite{zaba})
\be
\rho _* (\vec{u} )= \left({\frac{\beta}{2\pi q}}\right)^{3/2}\,  \exp \left(- {\frac{\beta \vec{u}^2}{2q}}\right).
\ee
It is a gaussian  with mean zero and variance $q/\beta $, devoid of any   rotational features.
The standard fluctuation-dissipation relationship can be retrieved by  introducing  the notation $D=q/\beta^2$, with $q= k_BT\beta/m$,  \cite{czopnik}.

\subsection{Disregarding magnetism.}

  Let us keep intact  the friction term in Eq. (42), but completely disregard  the magnetic contribution.  Setting $\vec{B}= \vec{0}$, we pass to  $\vec{F}= - \beta \vec{u}$, and thence arrive at   the familiar   Ornstein-Uhlenbeck process. Its  transition pdf  comes out in   the canonical form:
\be
p(\vec{v_0},0,\vec{u},t)  = \left( 2\pi \beta D t \right)^{- 3/2}  \,
 \exp \left( -  {\frac{(\vec{u} - e^{-\beta t}\vec{v_0})^2}{2 \beta D(1- e^{-2\beta t})}}\right).
\ee
The OU process relaxes to  the asymptotic pdf of  the form  (47) with $q/\beta ^2 = \beta D$.

The corresponding  diffusion generator  (in  velocity space)  is
\be
L= \beta ^2 D \Delta _{\vec{u}} -   \beta \vec{u}\cdot \vec{\nabla }_{\vec{u}},
\ee
while the Fokker-Planck  operator reads
\be
L^*=\beta ^2 D \Delta _{\vec{u}} +   \beta \vec{\nabla }_{\vec{u}}\cdot (\vec{u} \cdot ).
\ee
The process is time homogeneous, hence we can freely  replace  $t$ by  $(t-s)$ and  $\vec{v}_0$   by $\vec{v}$,   in the formula (48).

\section{Transformation to  spatial  non-conservative  processes.}

All previous derivations were intended to set the grounds for a clean  identification of spatial  magnetic-looking diffusion processes, which are actually   devoid of any obvious  phase-space connotations, c.f. \cite{czopnik}, while staying in conformity with the general formalism of Ref. \cite{monthus}, specified to  $N=3$.

Note that for a negatively charged particle $q_e  = - |q_e|$,  we have   $\omega_c= - \omega$, with $\omega = |\omega _c|$  and thence $\vec{F}(\vec{u}) =  \omega  (-u_2,u_1,0)$.
Not incidentally, vector functions of this  particular functional  form,   like  e.g.  $\vec{A} = (B/2) (-y,x,0)$,  notoriously   appear  in the discussion of the  minimal  electromagnetic coupling  in quantum theory, c.f. \cite{avron,roep}, and  likewise  in the description of  non-conservative   random dynamical systems, \cite{zambrini,gar,wiegel}.

  In view of   $\vec{\nabla } \times \vec{A}  = (0,0,B) = \vec{B} $, we may interpret $\vec{A}$  as an exemplary vector  potential of the  magnetic field   $\vec{B}$  oriented in the $z$-direction. We  thus arrive at another (alternative  to  this described in Refs. \cite{czopnik,czopnik1})  view on the  implementation of electromagnetic perturbations of random motion,  developed in  the   Euclidean-looking  random motion  theory of   Refs. \cite{zambrini,gar}.

All basic formulas of Section III can be easily transformed from the velocity space  diffusion  to  the  closely affine spatial one. Our aim is to establish a   direct link with the notation of Ref. \cite{monthus}.
This needs  $\vec{u} \to \vec{x}$ replacements, and proper  adjustments of various parameters.

\subsubsection{No friction.}
We shall rely on notational conventions of \cite{gar,avron}. Let us recast  Eq.  (1)  to the  form (we set $\nu =1/2$):
\be
d\vec{X}(t) = \vec{A}(\vec{X}(t)) dt +  d\vec{W}(t),
\ee
where  the   former  drift  $\vec{F}(\vec{x})$  is replaced by      $ \vec{A}(\vec{x})=  (B/2) (-x_2,  x_1, 0)$.
 Accordingly, the Fokker-Planck equation (2) takes  the form
\be
\partial _t \rho = {\frac{1}2}  \Delta \rho - \vec{\nabla } (\vec{A} \rho ),
\ee
with   $\vec{A}(\vec{x}(t))= + \Lambda \vec{x}(t)$, where
\be
\Lambda =
\left(
  \begin{array}{ccc}
    0 & -B/2 & 0 \\
    B/2 & 0 & 0 \\
    0 & 0 & 0 \\
  \end{array}
\right).
\ee
The resultant transition probability density function retains  the  functional form  (45), except for the obvious   replacements  $\vec{u} \to \vec{x}$  and   $\omega _c \to    -B/2 $.

Clearly:
 \be
p(\vec{y},s,\vec{x},t)  = \left( {\frac{1}{2\pi (t-s)}}\right)^{3/2}  \, \exp \left[  -  {\frac{(\vec{x} - U(t-s)\vec{y})^2}{2(t-s)}}\right],
\ee
with
\be
U(t) = \left(
         \begin{array}{ccc}
          cos(Bt/2) & -sin(Bt/2)&0\\
           +sin(Bt/2) & cos(Bt/2) &0\\
           0&0& 1\\
         \end{array}
       \right),
\ee
and $0\leq s<t$,  is a fundamental solution of Eq.(51).

The transition pdf (54) solves the pair of equations  (3) and (6) with adjoint  generators
\be
\begin{split}
L^* ={\frac{1}2} \Delta - \vec{A}\vec{\nabla} =  -
\left[-{\frac{1}2}(\vec{\nabla }- \vec{A})^2 + {\cal{V}}\right],\\
 L ={\frac{1}2} \Delta + \vec{A}\vec{\nabla} = -
\left[-{\frac{1}2}(\vec{\nabla } + \vec{A})^2  +  {\cal{V}}\right],\\
\end{split}
\ee
where in view of $\vec{\nabla } \cdot \vec{A} =0$, we have   ${\cal{V}} = \vec{A}^2= (B^2/4)(x^2 + y^2)$.\\

We realize, that the Lagrangian (14)  takes here the form  (29), but   without the $V$ contribution, i.e.
\be
 {\cal{L}} = {\frac{1}2} \left[ \dot{\vec{x}} - \vec{A}(\vec{x}(\tau ))\right]^2.
\ee

\subsubsection{Frictional case.}

 The frictional case of Section IV.B gets transformed accordingly.  If we additionally set $\beta =1$, we arrive at
 \be
p(\vec{y},s,\vec{x},t)  = \left[ \pi (1- e^{-2 (t-s)})\right]^{- 3/2}  \,
 \exp \left( - {\frac{(\vec{x} - e^{-(t-s)} U(t-s)\vec{y})^2}{(1- e^{-2(t-s)})}}\right).
\ee
However, now the drift term in the Fokker-Planck equation (1) is a sum of two contributions:
\be
\vec{F} = \vec{A} -  \vec{x} = \vec{A} - \vec{\nabla } \phi  ,
\ee
 where $\vec{A}= (B/2)(-y,x,0)$,  $\vec{x}= (x,y,z)$  and   $\phi  = \vec{x}^2/2$.
  We realize that  $\vec{A}= (1/2) (\vec{B} \times \vec{x})$, with $\vec{B}= B(0,0,1)$, and thus  $\vec{\nabla}\cdot \vec{F}= - 3$.

The Fokker-Planck operator   $L^*$ appears in the functional  form  of Eq. (7), with the scalar potential (8):
\be
\begin{split}
&L^*= {\frac{1}2}\Delta - \vec{F}\cdot \vec{\nabla } + 3 = -[-{\frac{1}2}(\vec{\nabla} - \vec{F})^2 + {\cal{V}}],\\
&{\cal{V}} = {\frac{1}2} (\vec{F}^2   - 3) ={\frac{\vec{A}^2}2} +  {\frac{1}2}(\vec{x}^2  - 3).\\
\end{split}
\ee
It is perhaps amusing to note that ${\cal{V}}$ vanishes on the ellipsoid  $\vec{A}^2 + \vec{x}^2 = 3$,
while taking negative values in its interior and positive on the outer side.

At this point,  we can resort to the discussion of Section II, where the condition (12) validates the coexistence of the steady current with the  asymptotic  stationary pdf.
Indeed, the stationary pdf derives from:
\be
\rho _*(\vec{x}) = \pi  ^{-3/2}\,  \exp  (-\vec{x}^2),
\ee
  Since   $\vec{A}= (B/2) (-y,x,0)$,  we have  $\vec{\nabla }\cdot \vec{A} = 0$,  in conjunction with
  $\phi (\vec{x}) = \vec{x}^2 /2$. Accordingly  the condition (12) is valid, and the diffusion process (57) admits the divergenceless steady current
  \be
   j_*(\vec{x}) = \vec{A}(\vec{x})\,    \rho _* (\vec{x}).
  \ee

The present case actually is an  explicit  illustration  to our discussion of Section III.B,  concerning  Lagrangian signatures of stationary pdfs,  and the existence of steady currents. The corresponding Lagrangian  stems from the general  formula (15). However, in view of Eq. (25) it  can be reduced to the effective  form (29):
$
{\cal{L}}_{magn}=
{\frac{1}2} \left[ \dot{\vec{x}}(\tau ) - \vec{A}(\vec{x}(\tau ))\right]^2  + V(\vec{x}(\tau)).
$
which entails the evaluation of the integral kernel (27). The  F-P pdf (26) emerges after accounting for the multiplicative conditioning of this kernel.

\subsubsection{No magnetism.}

By disregarding the magnetic contribution, we arrive at   a simplified  form of the transition function  (58), which actually is  the transition pdf of the Ornstein-Uhlenbeck process with the drift $\vec{F} = -  \vec{x} = -\vec{\nabla }\phi $, where  $\phi =  \vec{x}^2/2$, c.f. also \cite{gar3,zaba}, (here $0\leq  s < t$):
\be
p(\vec{y},s,\vec{x},t)  = \left[ \pi  (t-s) \right]^{- 3/2}  \,
 \exp \left( -  {\frac{(\vec{x} - e^{-(t-s)}\vec{y})^2}{(1- e^{-2(t-s)})}}\right).
\ee
This transition pdf is intimately intertwined  with the  integral kernel of $\exp(-tH)$, where $H$ is the quantum harmonic oscillator Hamiltonian (with properly tuned parameters), \cite{avron,zaba}.

Namely, we have
\be
p(\vec{y},s,\vec{x},t) = e^{3(t-s)/2}\,  k(\vec{y},s,\vec{x},t) {\frac{\phi _1(\vec{x})}{\phi _1(\vec{y})}},
\ee
where $\Phi _1(\vec{x})= \pi ^{-3/2}\exp(- \vec{x}^2)$,  is the ground state function,  while the factor  $3/2$  in the exponent  is  the lowest  eigenvalue  of  $H= (1/2)(-\Delta   +  \vec{x}^2)$. The function   $k(\vec{y},s,\vec{x},t)$ is the  integral kernel of $\exp[-(t-s)H]$, see e.g. Section III.B, where $H_{st}$ differs from $H$ by an additive renormalisation $-3/2$.

 The kernel, whose evaluation belongs to the standard  path integral inventory  (here, in association with the  Feynman-Kac formula),  has the form, \cite{zaba,avron,roep}:
\be
\begin{split}
k(\vec{y}, \vec{x},t) =  \exp(-3t/2) (\pi [1- \exp(-2t)])^{-3/2} \exp \left[ {\frac{1}2} (\vec{x}^2- \vec{y}^2) - {\frac{(\vec{x} - e^{-t}\vec{y})^2}{(1-e^{-2t})}}\right],\\
=(2\pi \sinh t)^{-3/2} \ \exp\left[ - {\frac{(\vec{x}^2 + \vec{y}^2) \cosh t- 2\vec{x}\cdot \vec{y}}{2 \sinh t}}\right].\\
\end{split}
\ee
It is instructive to mention that the formal replacement  of  $t$ by  $it$  in Eq. (64), reproduces the familiar   propagator   $\exp(-itH)(\vec{y},\vec{x})$ of the quantum mechanical harmonic oscillator in $R^3$, \cite{roep}  (keeping in memory that we have  scaled away all physical constants).

We point out that the kernel  (65)  of the semigroup $\exp(-tH)$   has the Feynman-Kac  path integral representation with the Lagrangian     ${\cal{L}}=  {\frac{1}2}[\dot{\vec{x}}^2(\tau ) +  \vec{x}^2(\tau )]$, compare e.g. (32).\\

Coming back to the OU  probability density function (63),  we may recall the   arguments of Section III.B, while carried out with $\vec{A}$ skipped in all formulas.  The corresponding Lagrangian has the  form:
\be
{\cal{L}}={\frac{1}2} (\dot{\vec{x}} - \vec{F})^2  +  \vec{\nabla }\cdot \vec{F} =  {\frac{1}2}\dot{\vec{x}}^2(\tau ) +  \dot{\vec{x}}\cdot \vec{x} +   V(\vec{x}(\tau )),
\ee
with $V = (1/2)\vec{x}^2 - 3/2$  and  $\vec{F}=- \vec{x}$.  The Lagrangian (66) rewrites as ${\cal{L}}= (1/2)(\dot{\vec{x}} + \vec{x})^2  - 3$, compare e.g. \cite{grosche}, chap. 6.2.1.17.

The  related  Fokker-Planck operator  derives directly from (60) by skipping the $\vec{A}^2$ term. Accordingly  $L^* ={\frac{1}2}\Delta + \vec{x}\cdot \vec{\nabla } + 3 = -[-{\frac{1}2}(\vec{\nabla} + \vec{x})^2 + V]$.  We recall that the standard form  of $L^*  = (1/2) \Delta  + \vec{\nabla }(\vec{x} \, \cdot )$  directly  follows  from Eq. (2).

\section{Probabilistic  significance of  path integral kernels. Obstacles to be observed. }

 All analytic formulas for transition probability densities  of $D=3$ diffusion processes discussed in Section V, can be  obtained  (independently  form our reasoning based on \cite{czopnik,czopnik1})  by means of standard path integration arguments,  \cite{wiegel,hunt,grosche}. The evaluation   often   becomes  quite tedious and  computationally  sophisticated, c.f. comments following Eq. (22) in Ref. \cite{wiegel}.

  Although we have not evaluated explicitly  any of  these integrals in the present paper,  in  each subsection of Section V we have reproduced the form of the  appropriate  Lagrangian,  which can be employed  in the general  path integral  formula (15). In this connection see e.g.   Appendix D  in Ref. \cite{monthus} for a discussion of $N\geq 2$ path integral procedures for quadratic Lagrangians. The complementary discussion for   $N=1$ can be found in \cite{pell}.

Since  the  Euclidean map of Section I.A is quantally motivated, we need to remember about  the wealth of available path integral formulas for propagators of various quantum systems, \cite{grosche}. One should as well be aware of  serious   jeopardies  threatening an  uncritical  usage of that map.  In the context of diffusion processes, all path integral kernels  of interest (not only the transition pdfs, c.f. Eqs.(26) and (31) for  hints)   {\it must}   {\it  necessarily  need to  be positive and secure semigroup properties}  (akin to  the Chapman-Kolmogorov identity  for Markovian pdfs).

 One should become alert by the fact, that   transition probability densities for diffusions with  non-conservative drifts,  of Sections I to V,  appear  as  (fundamental) solutions of the  Fokker-Planck equation with the generator $L^*=  - (H_{Eucl} +{\cal{V}})$.  Thus,  even for the simplest non-conservative drift  $\vec{F}= \vec{A}$ with $\vec{\nabla }\cdot \vec{A}=0$,  the operator $H_{Eucl}$  never appears alone as it stands, but  is always accompanied by  an additive correction  (perturbation)  $\vec{A}^2/2$.

  This raises doubts, \cite{gar},  about the probabilistic significance  (as far as  legitimate  diffusion processes are concerned)  of the  integral kernel of  the operator $\exp(-tH_{Eucl})$.      We point out that the  path integral  kernel  of $exp(- t H_{quant})$, with $t\geq 0$,   and $H_{quant} =-{\frac{1}2} (\vec{\nabla } - i\vec{A})^2$ is complex-valued, \cite{glimm,glasser,pell,grosche,gar}. The pertinent complex-valuedness can be removed by the supplementary Euclidean map $\vec{A} \rightarrow i\vec{A}$, \cite{zambrini}, or   $\vec{A} \rightarrow i\vec{A}$,  (to keep the convention of Ref. \cite{mazzolo}), c.f.  the Appendix.

\subsection{Link with Schr\"{o}dinger semigroups: ${\cal{L}}_{cl} \rightarrow  {\cal{L}}_{Wick}$.}

In the light of  our preliminary discussion in Section I.A, concerning the Euclidean map $it \rightarrow t, t \geq 0 $ (actually  the Wick rotation $\tau = it$),  it appears    useful to recall some features of this  formal affinity  between quantum mechanical and diffusion-type patterns of dynamical behavior (all physical constants are scaled away)  in the  simplest  case of  Hermitian Hamiltonians, when  Schr\"{o}dinger semigroups can be safely  introduced,  \cite{zaba,grosche}.

Given $H= -(1/2)\Delta +V$ with a confining potential $V$, we  execute: $
  \exp(-iHt) \psi _0 = \psi _t  \Longrightarrow
\exp(-tH)\Psi _0 = \Psi _t$.

In the absence of external potentials   (free case) we have  $H =  - (1/2)\Delta $ (keep in memory our  scaling  $\nu \rightarrow 1/2$.)
One knows that the familiar heat kernel
\be
p(\vec{y},\vec{x},t) \equiv   k(\vec{y},\vec{x},t)= [\exp({\frac{t}2}\Delta )](\vec{y},\vec{x}) = (2\pi t)^{-3/2}\, \exp[-(\vec{x}- \vec{y})^2/2t].
\ee
has a path integral representation (14) with the Lagrangian  (15) reduced to  ${\cal{L}}  = (1/2)\dot{\vec{x}}^2(\tau )$  (i.e.  ${\cal{L}}_{Wick}$ of Section I.A).

 This kernel can be  formally   deduced by  performing   $t\rightarrow it$  in the  free Schr\"{o}dinger   propagator
\be
K(\vec{y},\vec{x},t)= [\exp(i{\frac{t}2} \Delta )](\vec{y},\vec{x}) = (2\pi it)^{-1/2}\, \exp[+i(\vec{x}-\vec{y})^2/2t],
\ee
which  is associated   with ${\cal{L}}_{cl}= (1/2)\dot{\vec{q}}^2(t)$, c.f. Section I.A., via
the  Feynman  path integrand $\exp[i\int_0^t  ds\, {\cal{L}}_{cl}(\dot{\vec{q}}(s),\vec{q}(s)]$, \cite{roep,grosche}.

Let us consider $H= (1/2)(-\Delta + \vec{x}^2 ) $,  which differs from $H_{st}$  of Section III.B   by the missing  additive renormalisation  constant $-3/2$, c.f.   (63)-(66).
The integral kernel  $k(\vec{y},\vec{x},t)$ of $\exp(-tH)$  is given by   Eq. (65), provided we replace $(t-s)$ by $t$.  The (Euclidean, actually Wick) Lagrangian in the  path integral (Feynman-Kac) formula for   (65) reads  ${\cal{L}}_{Wick} =  (1/2)[\dot{\vec{x}}^2(\tau )  + \vec{x}^2(\tau )]$.

By formally executing  $t\rightarrow it$  in the formula (65), one  arrives at the   quantum harmonic oscillator  propagator
\begin{equation}
 K(\vec{y},\vec{x},t) = [\exp(-  iHt)] (\vec{y},\vec{x}) = (2\pi i \sin t)^{-3/2}   \exp  \left[ +i {\frac{(\vec{x}^2+\vec{y}^2)\cos t  - 2\vec{x}\cdot \vec{y}}{2\sin t }} \right],
 \end{equation}
associated with  ${\cal{L}}_{cl} =  (1/2)[\dot{\vec{q}}^2(t) - \vec{q}^2(t)]$.

Note that we here bypass the problem of giving meaning to square roots of trigonometric functions when they take negative values, e.g. an issue of Maslov corrections (indices),  \cite{grosche}.\\

Eq. (69) should be set against its semigroup version (c.f (65))
\be
k(\vec{y}, \vec{x},t) = \exp(- tH)(\vec{y},\vec{x})= (2\pi \sinh t)^{-3/2} \ \exp\left[ - {\frac{(\vec{x}^2 + \vec{y}^2) \cosh t- 2\vec{x}\cdot \vec{y}}{2 \sinh t}}\right].
\ee
We note that the above semigroup kernel can be related to the legitimate transition  pdf of the OU process (63), by means of the  conditioning formula (64), see  also (26).\\

{\bf Remark 5:} In relation to  the "Euclidean  time" label, we  may invoke  the statistical physics lore of the $50$-ies and $60$-ties, by  passing   to an integral kernel of the density
operator,  that is parameterized  by equilibrium values of the temperature. To this end one should  set e.g.  $t\equiv \hbar\omega /k_BT$ for a  harmonic  oscillator with a proper   frequency $\omega $ and  remember about evaluating the normalization factor $1/Z_T$,  where $Z_T$ stands for a partition function of the system, \cite{roep}. In particular, let us  mention that   the   complex-valued integral kernel of  $exp(-\beta H_{quant})$  with $\beta \sim 1/k_BT$, where $T$ stands for a temperature, can be interpreted   as an  unnormalized density matrix in the study  the  diamagnetism  of free electrons, \cite{sond,glasser}.
\\

\subsection{Charged particle in a constant vector potential.}

We shall look for  possible deficits  of the  formal  map ${\cal{L}}_{cl} \rightarrow  {\cal{L}}_{Wick}   \rightarrow {\cal{L}}_{Eucl}$,  by invoking   a  simple  example of the  Schr\"{o}dinger  propagator for a  "free particle with a (constant) vector potential", \cite{grosche}, chap. 6.2.1.3.  While adopted to our  notation  (with scaled away physical constants, presuming  the charge $e= +|e|\equiv 1$),  we consider  the motion operator  $\exp(-itH_{quant})$ with $H_{quant} = - (1/2)(\vec{\nabla } -i\vec{A})^2$, where $\vec{A}$ is a constant vector field.

 The   path integral formula  for  the integral kernel in question
\be
K(\vec{y},\vec{x},t)  =  \int {\cal{D}}(s)\exp[i\int_0^t  ds\, {\cal{L}}_{cl}(\dot{\vec{q}}(s),\vec{q}(s))]  =
(2\pi it)^{-3/2} \exp [i \frac{(\vec{x}- \vec{y})^2}{2t}) + i\vec{A} \cdot (\vec{x}-\vec{y})]
\ee
derives from  the Lagrangian
\be
{\cal{L}}_{cl}=
(1/2) \dot{\vec{q}}^2(t)  +  \dot{\vec{q}}(t) \cdot  \vec{A}.
\ee

The Euclidean map ${\cal{L}}_{cl} \rightarrow {\cal{L}}_{Wick}$ leads to the integral kernel  of $\exp(-\tau H_{quant})$,   deriving from the Lagrangian
\be
{\cal{L}}_{Wick}=
(1/2) \dot{\vec{x}}^2(\tau )  -  i \dot{\vec{q}}(\tau ) \cdot  \vec{A}.
\ee

The subsequent map $\vec{A} \rightarrow -i\vec{A}$   defines  na incomplete Lagrangian (15), with the (constant)  term $\vec{A}^2/2$  actually  missing
\be
{\cal{L}}^{inc}_{Eucl}(\vec{x}(\tau ), \dot{\vec{x}}(\tau )) = {\frac{1}2}\dot{\vec{x}}^2(\tau )  -  \dot{\vec{x}}(\tau )\cdot \vec{A}(\vec{x}( \tau )),
\ee
and  $K(\vec{y},\vec{x},t)$  gets  transformed into
\be
k^{inc}(\vec{y},\vec{x},t) =
(2\pi t)^{-3/2} \exp \left[-  \frac{(\vec{x}- \vec{y})^2}{2t} - \vec{A} \cdot (\vec{x}-\vec{y})\right].
\ee
The "incompleteness" of both expressions  (73) and (74)  can in principle be  healed. A clean parallel with formulas (14) and (15)  may be established  by  incorporating   an additive correction (perturbation)
\be
{\cal{L}}_{Eucl} ={\cal{L}}^{inc}_{Eucl} + (1/2)\vec{A}^2,
\ee
which, via (14) asserts that we recover a legitimate transition pdf
\be
p(\vec{y},\vec{x},t) =
(2\pi t)^{-3/2} \exp \left[-  \frac{(\vec{x}- \vec{y})^2}{2t} - \vec{A} \cdot (\vec{x}-\vec{y})  - t{\frac{\vec{A}^2}2}\right]= (2\pi t)^{-3/2} \exp\left\{- {\frac{[(\vec{x}- \vec{y}) +  t \vec{A}]^2}{2t}} \right\}
\ee
of the   fairly standard version  of the diffusion  process   (1)   with a  {\it constant} vector drift  field  $\vec{F} \equiv \vec{A}$.

Indeed, the appearance of the  additional term $\vec{A}^2/2$ in  (75), amounts to multiplying the  kernel (74) by $\exp[- (t/2)\vec{A}^2]$. We point out that the diffusion process  pdf (76) is   consistently associated with the integral kernel (74) due to: (i) positive-definiteness of the kernel, (ii)  a  taming  factor  $\exp(-t\vec{A}^2/2)$  turning $k$ into $p$ according to $p(\vec{y},\vec{x},t) =   \exp(-t\vec{A}^2) k^{inc}(\vec{y},\vec{x},t)$.

\subsection{Solenoidal vector  potential.}

Presently, we shall discuss relationships (maps) between path integral kernels (propagators) dderiving from Lagrangians (72)-(74), under the assumption the $\vec{A}$ is no longer a constant vector, but a solenoidal  vector  field (devoid of any explicit time-dependence).
Our choice is $\vec{A}= (B/2)(-y,x,0)$ so that $\vec{\nabla } \times \vec{A}= (0,0,B)$. We shall assume $B=2$ at some point in below.

The path integration with the Lagrangian of the form (72) in the action functional, is a text-book classic (we recall about scaled away physical constants, including the pre-definiton of charges  which equate  $ \pm 1$). The propagator reads:
\be
\begin{split}
& \exp (-itH_{quant})(\vec{y}, \vec{x}) = \left({\frac{1}{2\pi it}}\right)^{3/2}
 \left({\frac{B/2} {\sin(Bt/2)}}\right) \cdot  \\
& \exp\left\{ (i/2)\left[ B(-x_1y_2+x_2y_1)+[(y_1-x_1)^2+(y_2-x_2)^2]{\frac{B/2}{\tan (Bt/2)}} +{\frac{(y_3-x_3)^2}t}\right] \right\}.\\
\end{split}
\ee
with $H_{quant} = - (1/2) (\vec{\nabla } - i\vec{A})^2$.

This may be directly compared with standard Feynman's  path integral  formula for the charged particle in a uniform magnetic field, while setting  $m=1$, $\hbar=1=c$ and $\omega =B$ , \cite{roep,grosche}.

The semigroup kernel, given by the path integral formula stemming from he Lagrangian (73) is as well a literature classic, with the wealth of various evaluation procedures, \cite{avron,roep}. It reads:
\be
\begin{split}
& \exp (-tH_{quant})(\vec{y}, \vec{x}) = \frac{B}{4\pi \sinh(Bt/2)}\left(\frac{1}{2\pi t}\right)^{1/2} \cdot  \\
& \exp\left\{ (iB/2)(-x_1y_2+x_2y_1)-\frac{B}{4}\left[(y_1-x_1)^2+(y_2-x_2)^2\right]\coth (Bt/2) -\frac{(y_3-x_3)^2}{2t}\right\}.\\
\end{split}
\ee
Note that by setting $t\rightarrow it$ in Eq. (79) (this, in conformity with the map $\tau \equiv it$ of Section I.A),   we recover the formula (78).

The presence of an imaginary factor  $i$ in the first term of the exponent, makes the kernel complex-valued. This precludes its probabilistic significance  within the diffusion process   framework of sections I-V, were only real-valued and strictly positive kernel  functions were admitted.

By    {\it formally}  passing to the imaginary magnetic field $B\rightarrow -iB$  (strictly speaking $\vec{A} \rightarrow -i\vec{A})$,   we obtain (see also  \cite{gar})   the   real-valued kernel function
\be
\begin{split}
& \exp (-tH_{Eucl})(\vec{y}, \vec{x}) = \frac{B}{4\pi \sin(Bt/2)}\left(\frac{1}{2\pi t}\right)^{1/2} \cdot  \\
& \exp\left\{ - (B/2)(-x_1y_2+x_2y_1)-\frac{B}{4}\left[(y_1-x_1)^2+(y_2-x_2)^2\right]\cot (Bt/2) -\frac{(y_3-x_3)^2}{2t}\right\}.\\
\end{split}
\ee

However, an  obvious problem needs to be raised, \cite{gar}. Since trigonometric functions have replaced   the hyperbolic ones in the expression  (80),  the kernel function is real-valued, but  remains positive  {\it only}   for times   $0< tB/2\in (2n\pi,2n\pi + \pi /2)$ with $n=0, 1, 2,...$   \cite{gar}.  This time limitation  allows as well   to bypass  the   negative  sign   problem  for    $cot (Bt/2)$,  which would   make the kernel  an unbounded function  (an exemplary signature of this  misbehavior   is   $\cot (Bt/2) <0$ for $Bt/2\in (\pi /2, \pi )$,  approaching $- \infty $ for $Bt/2 \to \pi $).

The  above  sign/unboundedness obstacles  have been created by a formal transformation $\vec{A} \rightarrow -i\vec{A})$,  while  executed directly in the path integral kernel expression (79).
 To keep this  formal  transformation under   control,  we   provide  in the Appendix   a  detailed  evaluation of the path integral kernel $\exp (-tH_{Eucl})(\vec{y}, \vec{x})$, by  employing the   "canonical"  methodology,  suitable for quadratic Lagrangians, \cite{roep}.

\subsubsection{Generators versus Lagrangians. Resume.}

Taking seriously the path integral viewpoint of Section III,  we should try to verify, whether the assumption of the  diffusive motion can at  be reconciled  with the  dynamics induced by   motion   operators $e^{-tH_{Eucl}}$ and $e^{-tH^*_{Eucl}}$, see also Section VI.   In case of solenoidal vector field $\vec{A}$, $\vec{\nabla }\cdot \vec{A} =0$, we have:
\be
\begin{split}
H_{Eucl} =  - {\frac{1}2} (\vec{\nabla }  - \vec{A})^2  = - {\frac{1}2}\Delta + \vec{A}\cdot \vec{\nabla } - {\frac{1}2}\vec{A}^2 ,\\
H^*_{Eucl}   =  -  {\frac{1}2} (\vec{\nabla }  + \vec{A})^2 =    - {\frac{1}2}\Delta - \vec{A}\cdot \vec{\nabla } - {\frac{1}2}\vec{A}^2, \\
\end{split}
\ee
and recall  that  $H_{Eucl}$ and $H^*_{Eucl}$ are  Euclidean analogues of standard Hermitian operators, appropriate (up to dimensional scalings) for the  quantum Schr\"{o}dinger dynamics with the minimal electromagnetic coupling: $H^{\pm }_{quant} = - (1/2) (\vec{\nabla } \pm i\vec{A})^2 $.

In passing w note that     $ H_{Eucl} +  {\frac{1}2} \vec{A}^2 =  -  L^* $, c.f. Eq. (7), where $L^*$   is  the Fokker-Planck   generator of  a legitimate diffusion process with the forward drift  $\vec{F} = \vec{A}$. Compare e.g. Sections V.1 and VI.B, in particular Eqs. (56).

To evaluate the propagator of $\exp(- tH_{Eucl})(\vec{y}, \vec{x}) = k(\vec{y},0,\vec{x},t)$, with $H_{Eucl}=
 - {\frac{1}2} (\vec{\nabla }  - \vec{A})^2$,  we  refer to  (56), and  the   Lagrangian  (57), but presently
 without the  ${\cal{V}}$  contribution. Accordingly:
\be
\mathcal{L}_{Eucl}= \frac{\dot{\vec{x}}^2}{2}  - \dot{\vec{x}}\cdot\vec{A}.
\ee

For clarity of discussion, we refer to Eqs. (21), (22) in Section III, where the classical (non-Euclidean) Lagrangian for a charged particle in the electromagnetic field has been invoked, c.f. also \cite{cohen}..
The charge label  $q_c$  has been  left  undefined. Let us choose $q_c= - |q_c|= -1$, and additionally $m=1$.
Then $\mathcal{L}$ (22)  for a particle in a  magnetic field $B$  takes the form (we presume $U=0$)
\be
\mathcal{L}=\frac{\dot{\vec{x}}^2}{2} + \dot{\vec{x}}\cdot\vec{A}.
\ee
We note that  by  choosing  in (22)   the opposite (positive) sign of the charge $q_c=  |q_c|= 1$, we actually  would have changed  the sign of the second term in Eq. (83). (The same sign change would result from replacing
$\vec{A}$ by $- \vec{A}$).

Accordingly,   $\mathcal{L}=\frac{\dot{\vec{x}}^2}{2} -  \dot{\vec{x}}\cdot\vec{A}$,  (for positive charge)  formally  (set $m=1$)  coincides  with   $\mathcal{L}_{Eucl}$,(83) appropriate for   the negative charge  (c.f. our discussion about sign changes in the Euler-Lagrange equations in Section III).

\subsubsection{The (negative) outcome.}

The {\it resultant  kernel function}, whose  explicit evaluation has been transferred  to   the Appendix,  ultimately appears  in the form  "almost identical" with (80), see e.g (A.21)

\be
k(\vec{y},0,\vec{x},t)=\frac{1}{2\pi |\sin t|} \left(\frac{1}{2\pi t}\right)^{1/2}\exp\left\{-x_1y_2+x_2y_1-\frac{1}{2}\left[(y_1-x_1)^2+(y_2-x_2)^2\right]\cot t-\frac{(y_3-x_3)^2}{2t}\right\}.
\ee

The kernel is real-valued, positive, but unbounded (that in view of the sign changes of the $\cot t$ factor in the Gaussian exponent). Its another serious deficit is that the semigroup composition law (akin to the Chapman-Kolmogorow identity) is not valid for all times $t \in R^+$, see e.g (A.27).

We point out that  the    $1/sin(Bt/2)$  of (80),  in (A.21)  has been replaced by the $1/|sin(t/2)|$,  (we indicate that  the  pertinent derivation refers to a simplified case of $B=2$).
 An  explicit presence of  the  $\cot t$ multiplier in the exponent of the Gaussian function in (A.21), prohibits the validity of the semigroup composition law (see Section A.), unless $\cot t >0$, i.e. for time intervals $t\in (2n\pi , 2n\pi +\pi /2)$ only.

  We point out that contrary to the wealth of  evaluation procedures, which  are known  for  the integral kernel  of $\exp(-tH_{quant})$, \cite{avron,roep}, to our knowledge no  explicit evaluation of the kernel for $\exp(-tH_{Eucl})$ has been  available in  the literature.  The derivation of (A.21)   stems   directly from the   path integration  basic principles,  \cite{roep,grosche}. To the contrary,  the expression (80), has been  obtained    by means of  a formal analytic continuation  procedure executed  directly  in   the  function $\exp(-tH_{quant})(\vec{y}, \vec{x})$, (79).

The conclusion is,  that the path integration procedure  can be completed   for  $\exp[-t H_{Eucl}] (\vec{x}, \vec{y})$   as a matter of principle. That is  regarded as the  action functional evaluation  problem  for quadratic Lagrangians.   However, the outcome (i.e.  kernel functions  (80) and  (A.21)),  is  incompatible  with the presumed     Markovian diffusion picture.   Clearly,   the motion operator    $\exp[-t H_{Eucl}]$  cannot be  "as it stands"  related to any legitimate diffusion process within the ramifications of sections I-V.    This defect can be overcome by    admitting   perturbations by   suitable  scalar potentials.   This  was the case in our  previous  discussion of   solutions of the Fokker-Planck equation with the generator $L^* = - ( H_{Eucl} + {\cal{V}})$ in Sections V and VI.B.1. where the (indispensable to secure positivity, continuity  and boundedness properties)   minimal additive perturbation had the form $\vec{A}^2/2$. \\

{\bf Remark 6:}
 We  mention that the kernel  formula for $\exp(-tH_{Eucl})$, displayed   in  Ref. \cite{zambrini}, sect. 5.2, p 92, is  incorrect.   In addition to previous outcomes (78)-(80),  we have made a direct check  (tedious, with the Wolfram Mathematica 12 assistance) to demonstrate that  the  kernel  of  $\exp(-tH_{Eucl})$, Eq. (80), with $B=2$,  actually is a solution  of the partial differential equation
 $$
\partial _t k  = -H_{Eucl} k = –{\frac{1}2} \Delta k + \vec{A}\cdot \vec{\nabla } k -  {\frac{1}2}  \vec{A}^2 k =  - {\frac{1}2}  (\vec{\nabla } - \vec{A})^2 k.
$$
Provided, all differentiations are carried out with respect to $\vec{x}$, and the time domain includes only  intervals where $sin (t)$ and $cot (t )$   are positive,  additionally   avoiding    the  infinitesimal vicinity of troublesome points like e.g. $2n\pi + \pi /2$, $n=0,2,...$.  Our check is  unquestionable  for $0<t< \pi /2$.

\section{Links with Euclidean quantum mechanics: Dissipative counterpart  of quantum dynamics  in $R^+$.}

\subsection{Agreeing the notation.}

  For clarity reasons, we refer to our  previous    discussion   (21)-(24), see also (34) -(42)  and introductory paragraphs of Section V, concerning an impact of physical dimensional constants, in particular that of the charge $q_c$ sign choice. The particular form  (39) of the vector field  $\vec{F}= \omega _c( u_2,-u_1,0)$  involves $\omega _c= q_cB/mc$ where $q_c$ stands for the charge label.
  Choosing the negative charge $q_c= - |q_c|$ we introduce  $\omega _c = - \omega, \omega >0$ and thence $\vec{F}= \omega (- u_2,+ u_1,0)$. After suitable rescaling, we may pass to the discussion of Section V, where $\vec{A}= (B/2)(-y,x,0)$ has encoded the negative charge sign input, and we have the notation fully agreed with that of sections I to V.

We recall that on physical grounds, \cite{cohen}, the  classical  Lagrangian for a charge in an electromagnetic field has the dimensional form
\be
{\cal{L}}_{cl}^{phys} =  {\frac{1}2} \dot{\vec{x}}^2 + \dot{\vec{x}}\cdot (q_c \vec{A})  - q_c U
\ee
where $\vec{A}$ is a  magnetic potential and $U$ is an electric  (scalar) potential.

To stay in conformity with the notation of Section I.A,   and our subsequent discussion of various Lagrangians,  we should  keep in mind that   $q_c \vec{A}\equiv \vec{F}$.  This observation is of relevance, once we pass to dimensionless expressions like   $L^* = - (H_{Eucl} + {\cal{V}})$ with    $H_{Eucl}= -{\frac{1}2} (\vec{\nabla }- \vec{F})^2$ and ${\cal{V}}= {\frac{1}2} (\vec{F}^2 + \vec{\nabla }\cdot \vec{F})$, c.f. (7)-(9).

We note that the standard dimensional form of the $H_{quant}$, \cite{cohen}, after rescaling  encodes $q_c \equiv \pm |q|$, where $|q|=1$:
\be
H_{quant}^{phys}  =  {\frac{1}{2m}} (-i\hbar \vec{\nabla }  - q_c\vec{A})^2  =
 - {\frac{\hbar ^2}{2m}} ( \vec{\nabla }- i {\frac{q_c}{\hbar}} \vec{A})^2  \rightarrow  H_{quant} = -{\frac{1}2}[\vec{\nabla } - i(\pm \vec{A})]^2,
\ee
 as  used throughout the paper.  We note that in Eq.(85), the sign choice for $q_c\vec{A}$ needs a parallel sign choice for $q_c U$.
Accordingly, choosing $\vec{F}\equiv \vec{A}$ we actually make an implicit choice of $q_c=+1$ in $q_cU$ as well.

\subsection{Further uses of Euclidean maps.}
\subsubsection{No magnetism.}

The rationale for Euclidean maps has been outlined in Section I.A.  From now on, instead of  introducing  $\tau = it$ for $t \in R^+$  to map ${\cal{L}}_{cl}$ into  ${\cal{L}}_{Wick}$, we shall  directly invoke a replacement   $t\rightarrow  -it$.  This results in $\exp(-iHt) \rightarrow \exp(-Ht)$, in conformity  with $it \rightarrow t$, actually  used in the preamble to Section VI.

Let us consider   the formal map $t \rightarrow -it$ on the level of (properly rescaled) Schr\"{o}dinger equation and its complex adjoint ( $\psi $ stands for a wave function,   while $\overline{\psi }$ for its complex conjugate).
Denoting  $H=-{\frac{1}2} \Delta  + \Omega $   the $L^2$ Hermitian generator of motion,  with virtually any confining potential $\Omega $ (this notation  is introduced for  the scalar potential, to avoid confusion  with previously employed  ${\cal{V}}$,  $V$ and $U$, purpose-dependent potentials), we adopt  a formal recipe:
\be
\begin{split}
i\partial _t  \psi  =  H \psi  \,\,  \rightarrow  \, \, \partial_t \theta _* = - H\theta _* , \\
 -i\partial _t \overline{\psi }  = H \overline{\psi } \, \, \rightarrow  \, \,  \partial_t \theta = H \theta .\\
\end{split}
\ee
C.f.  for rationale  in Refs. \cite{zambrini}-\cite{yasue} and  \cite{olk}-\cite{gar4}. All dynamical rules are confined to $t\geq 0$.

In terms of the Madelung (polar)  decomposition of Schr\"{o}dinger wave functions  (we implicitly
 presume  the $L^2$ normalization of $\psi $, so that $|\psi |^2 \equiv \rho$ has an interpretation of the probability density function)  we have:
\be
\begin{split}
\psi = \exp (R+ iS)  \rightarrow \theta _* = \exp (\overline{R} - \overline{S}),\\
\overline{\psi }  = exp(R-iS)  \rightarrow  \theta = \exp (\overline{R} + \overline{S}),\\
\end{split}
\ee
where $R, S, \overline{R}, \overline{S}$ are (appropriate) real-valued functions, while  the complex function  $\overline{\psi}$ is a  conjugate of  $\psi $.
The rationale (e.g. a more detailed relationship between $R, S$ and $\overline{R}, \overline {S}$) can be found in Refs. \cite{zambrini1,zambrini2}.

In view of (88), we have ($L^2$ normalization being implicit) $\overline{R}={\frac{1}2}  \ln (\theta _* \theta)$,  and   $\overline{S}=
{\frac{1}2}  \ln (\theta /\theta_*)$. We hereby anticipate that $\rho \rightarrow \overline{\rho } = \theta_* \theta $.

Except  for the overline gained by $R$ and $S$  in the mapping  to $\overline{R}$ and $\overline{S}$, the only
really significant  step  in  (83)    is the  Euclidean  mapping  accompanying $R\rightarrow \overline{R}$ (and thus $\rho \rightarrow \overline{\rho })$:
\be
 S \rightarrow i\overline{S}.
\ee

Let us check how the map (89)  works, if employed {\it formally}  in  the coupled system (of local conservation laws, \cite{czopnik,czopnik1}), which  derives directly from the Schr\"{o}dinger equation, \cite{zambrini1,zambrini2,nelson0,nelson}.  The  Hermitian generator choice  $ H=-{\frac{1}2} \Delta  + \Omega $, implies:
\be
\begin{split}
\partial _t R = - {\frac{1}2} \Delta S - \vec{\nabla}R \cdot \vec{\nabla} S,\\
\partial_t S = -  {\frac{1}2} (\vec{\nabla}S)^2  +   {\frac{1}2} \Delta R  + {\frac{1}2} (\vec{\nabla}R)^2 - \Omega .\\
\end{split}
\ee
We point out that the first equation (90) is equivalent to the continuity equation $\partial_t \rho =
 - \vec{\nabla }\cdot(\vec{v} \rho )$,  for $L$-normalised  $\rho = \psi  \overline{\psi } = |\psi |^2$, with the current velocity  $\vec{v} = \vec{\nabla } S$ in the gradient form, \cite{nelson0}.

By  setting in Eqs. (90) $R\rightarrow \overline{R}$, $t\rightarrow -it$ and $S \rightarrow i \overline{S}$, we arrive at:
\be
\begin{split}
\partial_t \overline{R} = - {\frac{1}2} \Delta \overline{S} - \vec{\nabla}\overline{R} \cdot \vec{\nabla} \overline{S},\\
\partial_t \overline{S} = -  {\frac{1}2} (\vec{\nabla}\overline{S})^2 -  {\frac{1}2} \Delta \overline{R}  - {\frac{1}2} (\vec{\nabla}\overline{R})^2   + \Omega .\\
\end{split}
\ee
The first equation is again a  continuity equation  $\partial_t \overline{\rho } = - \vec{\nabla }\cdot (\overline{V} \rho )$  for $L$-normalized  $\overline{\rho }= \theta \theta ^* $,  with $\overline{V}= \vec{\nabla }\overline{S}$ in the gradient form.

We note a conspicuous change of the sign of the potential terms, while passing from (90) to (91), compare e.g. also \cite{zambrini1,zambrini2,gar,zaba,olk,gar2}:
\be
\begin{split}
\partial_t S  + {\frac{1}2} (\vec{\nabla}S)^2  =  Q - \Omega ,\\
\partial_t \overline{S} + {\frac{1}2} (\vec{\nabla}\overline{S})^2 = \Omega  - \overline{Q}, \\
Q= Q(R)=  {\frac{1}2} (\Delta R  + (\vec{\nabla}R)^2)  \, \rightarrow  \overline{Q}= Q(\overline{R}),\\
\end{split}
\ee

In addition to current velocity fields   $\vec{v}= \vec{\nabla }S$  and $\vec{V}= \vec{\nabla } \overline{S}$, we introduce so-called osmotic velocities  $\vec{u} = \vec{\nabla }R = \vec{\nabla } \ln \rho ^{1/2}$, and anlgously  $\vec{U}$.
 By applying  the gradient to both sides of the dynamical laws in  (92), we get the familiar hydrodynamical evolution equations   (to be considered in conjunction with the corresponding  continuity equations), which  derive   respectively  from  the quantum and dissipative (here we anticipate an implicit  diffusive scenario) dynamics, \cite{zambrini1,hasegawa}:
\be
\begin{split}
\partial_t \vec{v}   +   (\vec{v}\cdot \vec{\nabla})\vec{v} = \vec{\nabla }(Q - \Omega),\\
 \partial_t \vec{V}   +   (\vec{V}\cdot \vec{\nabla})\vec{V} = \vec{\nabla }(\Omega - \overline{Q}).
\end{split}
\ee
We point out that in the derivation, the property   ${\frac{1}2} \vec{\nabla } \vec{v}^2 =   (\vec{v}\cdot \vec{\nabla})\vec{v}$, valid for gradient vector fields, has been employed.

We can give the specific potential $Q$ (de Broglie-Bohm quantum potential-related, \cite{gar4}) other equivalent forms. Namely, by setting $\vec{u} = {\frac{1}2} \vec{\nabla } \ln \rho $, we have
\be
Q = Q(\rho ) = {\frac{1}2} (\vec{u}^2 + \vec{\nabla} \cdot \vec{u}) = {\frac{1}2}  {\frac{\Delta \rho ^{1/2}}{\rho ^{1/2}}}.
\ee
See e.g. also  Ref. \cite{gar4}. An analogous formula holds true for $\overline{Q}= Q(\overline{\rho })$, with $\vec{u} \rightarrow \vec{U}= \vec{\nabla } \overline{R}$.

\subsubsection{With magnetism.}

Let us consider the rescaled Schr\"{o}dinger equation with the minimal magnetic coupling, \cite{cohen,nelson}, paired with its adjoint one (here $H_{quant} $  still is a Hermitian operator)
\be
\begin{split}
i \partial_t \psi  = [- {\frac{1}2} (\vec{\nabla } - i\vec{A})^2 + \Omega] = H_{quant}\psi , \\
-i\partial \overline{\psi } = H_{quant} \overline{\psi }. \\
\end{split}
\ee
Employing the Madelung decomposition (88), we arrive at the magnetic extension of the formulas (90), \cite{nelson0,nelson}:
\be
\begin{split}
\partial _t R =  - {\frac{1}2} \Delta S  - \vec{\nabla }R \cdot (\vec{\nabla }S -\vec{A}) + {\frac{1}2} \vec{\nabla } \cdot \vec{A},\\
 \partial _t S = -{\frac{1}2}(\vec{\nabla}S - \vec{A})^2 + Q - \Omega . \\
\end{split}
\ee
with $Q= Q(R)$ defined  in (92).    We introduce the current velocity in the non-gradient form (c.f. also \cite{nelson0})
\be
\vec{v} = \vec{\nabla } S - \vec {A}
\ee
and take the gradient of the second equation (96).
Since
\be
\begin{split}
{\frac{1}2}\nabla \vec{v}^2 =  \vec{v} \times (\vec{\nabla} \times \vec{v})  + \vec{v} \cdot \vec{\nabla }\vec{v},\\
\vec{\nabla } \times (\vec{v}
+ \vec{A}) = \vec{\nabla } \times \vec{\nabla }S=  0  \rightarrow \vec{\nabla }\times \vec{v} = - \vec{\nabla } \times \vec{A},\\
\end{split}
\ee
after setting  $\vec{\nabla } \partial _t S =  \partial _t \vec{V} +  \partial_t\vec{A}$  we   arrive at the local conservation law:
\be
\partial_t \vec{v}  + (\vec{v} \cdot \vec{\nabla })\vec{v}  =   \vec{v} \times (\vec{\nabla } \times \vec{A}) - \partial_t \vec{A}  +     \vec{\nabla }( Q - \Omega ).
\ee

With identifications $ \vec{B}= \vec{\nabla }\times \vec{A}$   and $\vec{E} =  -\partial_t\vec{A} - \vec{\nabla } \Omega $,  we
make explicit an impact of a minimal electromagnetic coupling on the level of the  inferred local conservation law:
\be
\partial_t \vec{v}  + (\vec{v} \cdot \vec{\nabla })\vec{v}  =  {\cal{F}}_{Lorentz}  +  \vec{\nabla } Q ,
\ee
with the Lorentz force acting upon the  charge $q_c=1$
\be
\begin{split}
{\cal{F}}_{Lorentz}  = \vec{E} +  \vec{v}  \times \vec{B}, \\
\vec{E} = - \partial _t \vec{A}  - \vec{\nabla } \Omega,\\
 \vec{B}=\vec{\nabla } \times \vec{A}.\\
\end{split}
\ee

We point out that the case of $\partial _t \vec{A}=0$, is of particular relevance  in connection with the discussion of Sections I to V, set e.g.  Eq. (17) against Eq. (24).

\subsubsection{Euclidean map in the presence of magnetism.}

Following the pattern of Section VII.B.2 we  should execute the maps $t\rightarrow -it$ and $S\rightarrow iS$.
Looking at the continuity equation (first equation (96)), we realise that to  be left with a real-valued expression, we must perform a  supplementary transformation $\vec{A} \rightarrow \pm i\vec{A}$.  In turn, the choice of $\vec{A} \rightarrow +  i\vec{A}$  would make incomplete  the  expected  sign inversion of the right-hand-side of Eq. (99).

Therefore,  we  opt to employ  the  $\vec{A} \rightarrow -  i\vec{A}$   convention  of Ref. \cite{monthus}, c.f. also Sections I to III in the present paper.
Accordingly, Eqs. (96) are mapped into:
\be
\begin{split}
\partial _t \overline{R} =  - {\frac{1}2} \Delta \overline{S}  - \vec{\nabla }\overline{R} \cdot (\vec{\nabla }\overline{S} +\vec{A}) - {\frac{1}2} \vec{\nabla } \cdot \vec{A},\\
 \partial _t \overline{S} = -{\frac{1}2}(\vec{\nabla}\overline{S} + \vec{A})^2 + \Omega  - Q . \\
\end{split}
\ee

Here, the formula (97) takes a Euclidean form  $\vec{V} = \vec{\nabla } \overline{S} + \vec {A} $, and   we readily recover
\be
\partial_t \vec{V}  + (\vec{V} \cdot \vec{\nabla })\vec{V}  =  -  [{\cal{F}}_{Lorentz}  +  \vec{\nabla } \overline{Q}],
\ee
in conjunction with $\partial_t \overline{\rho } = - \vec{\nabla}\cdot (\vec{V} \overline{\rho })$.  Note a  conspicuous change of sign of the force term in (103), if compared with (100).

Notice that our mappings actually imply $\vec{v} \rightarrow i\vec{V}$, which we may as well insert directly in  (99) to arrive at the outcome (102).

 Choosing the map $t\rightarrow -it $, $S\rightarrow iS$  and $\vec{A} \rightarrow -i\vec{A}$     in Schr\"{o}dinger equations (87) has profound  consequences.  The resultant Euclidean system  does not refer to Hermitian generators of motion, but to the non-Hermitian ones:
\be
\begin{split}
- \partial_t \theta _* = (H_{Eucl} + \Omega ) \theta  _*,\\
\partial_t \theta =  (H^*_{Eucl} + \Omega ) \theta ,\\
\end{split}
\ee
where we recall the factorisation ansatz     $\overline{\rho } =  \theta _* \theta $,  and  the definition of   $\overline{S}={\frac{1}2}  \ln (\theta /\theta_*)$, c.f. (88).\\

\subsubsection{Hasegawa's correspondence.}

 We realize that by merely identifying the external potential $\Omega $ with the previously  employed  ${\cal{V}}$, of Eqs. (8) and (12),   $\Omega \equiv {\cal{V}}$,  we uncover an obvious (albeit at the moment formal)  link with the diffusion processes  discussed at some length in Sections I to VI.
In particular, let us assume that actually  the forward drift of the Langevin-type equation (1)  equals  $\vec{F}= \vec{A}$. Then $\Omega = {\cal{V}} = {\frac{1}2}(\vec{A}^2 + \vec{\nabla }\cdot \vec{A})$.  The associated Fokker -Planck operator has the form
\be
L^* = -[H_{Eucl} +  {\cal{L} }]= - [ -{\frac{1}2}(\vec{\nabla } - \vec{A})^2 + {\frac{1}2}(\vec{A}^2 + \vec{\nabla }\cdot \vec{A})].
\ee

Clearly, we deal here with the special case of the system  (102), provided we identify $\theta_* $ with the solution of the Fokker-Planck equation $\overline{\rho }$, while $\theta $ is set identically equal $1$. This has been noticed  in Ref. \cite{hasegawa}, except for the wrong sign identification preceding $\Omega $ in  the analog of our formula (101)  (Eq. (9b) in \cite{hasegawa}.
Because  of $L={\frac{1}2}\Delta + \vec{F}\cdot \vec{\nabla} = - (H^*_{Eucl} + {\cal{V}})$,   a particular choice of $\theta =1$ provides  a legitimate solution of the equation $[H^*_{Eucl} + {\cal{V}}]\theta = \partial_t \theta $.

It has been noticed in  Ref.\cite{hasegawa},  that other  (e.g. without the $\theta =1 $ restriction) factorizations of the form   $\theta _* \theta = \overline{\rho }$  can be introduced, while in the framework of the coupled   system (103), with the potential $\Omega =  {\frac{1}2}(\vec{A}^2 + \vec{\nabla }\cdot \vec{A})$.  The corresponding current velocity in the continuity equation, then  acquires the form $\overline{V}=  \vec{A} - {\frac{1}2}\vec{\nabla } \ln (\theta / \theta^*)$ and refers to  conditioned diffusion processes (see our discussion in below).

\subsection{Schr\"{o}dinger's  boundary data and interpolation problem: Deciphering diffusive dynamics.}

Properly selected integral  kernels of various motion operators of the  form $\exp(-Ht)$, are vitally important  in connection with the concept of the Schr\"{o}dinger interpolating two-gate formula, and more generally in connection with the Schr\"{o}dinger boundary data problem (pertains to Schr\"{o}dinger bridges and Bernstein transition densities), \cite{schr,pavon,zambrini3,blanch,olk} see e.g. also \cite{zambrini,zambrini1,gar3,zaba}.

Integral kernels  $k(\vec{y},s,\vec{x},t)$  of a  probabilistic significance need to obey a number of restrictions, among which we list most  important for our purposes (see e.g. \cite{zambrini}, section 4):  strict positivity,   joint  continuity in all variables  $\vec{x}, \vec{y}, s,t$,  and semigroup composition property (analogue of the Chapman-Kolmogorov identity) $\int k(\vec{y},s,\vec{z},r) k(\vec{z},r,\vec{x},t) d^3z =  k(\vec{y},s,\vec{x},t)$, for $0\leq s<r<t\leq T$. (Not disregarding  the boundedness of the kernel fucntion.)

Given  $k(\vec{y},s,\vec{x},t)$  with the above  properties,   the  generated   time evolution we  consider in the   finite  time  interval $ [0,T]$ $0<s<t<T$. In principle $T>0$ can be arbitrarily large, and  $T\to  \infty $ limit may be often kept under control, c.f. \cite{gar3}, leading to well defined conditioned diffusion processes.

Note that the original formulation of the  Euclidean  quantum mechanics, has been time-symmetric from the outset and defined in the time interval $[-T/2,T/2]$, \cite{zambrini1,zambrini2}. Our subsequent discussion refers to the shifted time span $t \in [0,T]$   instead of $t\in [-T/2,T/2]$.

  We can  formally introduce  the (Bernstein, \cite{zambrini,zambrini1,hasegawa,gar3,olk})  probability density function with respect to $\vec{x}$ and $t$, with  a priori  fixed     initial $s=0, \vec{y}$ and terminal $r=T,\vec{z}$  variables:
\be
\rho _B(\vec{x},t)  = {\frac{k(\vec{y},0,\vec{x},t)\,  k(\vec{x},t,\vec{z},T)}{k(\vec{y},0,\vec{z},T)}}.
\ee
In view of the semigroup composition property, there holds  $\int \rho _B(\vec{x},t)d^3x = 1$ for all times $t\in [0,T]$.

 Given $\rho _B$, we may here  ask for a (Markovian, possibly  conditioned)  diffusion process  underlying   the time evolution of $\rho_B(\vec{x},t)$, \cite{zambrini,zambrini1,zambrini2,gar,gar3}.

Let us  introduce  a  more general version of Eq. (106), by resorting to functions $\theta{\vec(x},t)$ and $\theta ^*(\vec{y},s)$, which are obtained from  respectively terminal   $g(\vec{x})$,  and initial  $f(\vec{y})$ data for the evolution in the interval $[0,T]$:
\be
\begin{split}
\theta (\vec{x},t)= \int k(\vec{x},t,\vec{z},T)\, g(\vec{z})\, d^3z\\
\theta^*(\vec{x},t)= \int k(\vec{u},0,\vec{x},t)\,  f(\vec{u}) \, d^3u.\\
\end{split}
\ee

Let us anticipate that  (we no longer use the subscript $B$)
\be
\rho (\vec{x},t) = \theta^*(\vec{x},t) \, \theta (\vec{x},t)
\ee
actually stands for a  factorised  probability density function. Integrating with respect to $\vec{x}$ and admitting interchanges of involved integrals, we arrive at:
\be
\int \rho (\vec{x})\, d^3x = \int d^3u \int d^3z \,   f(\vec{u}) k(\vec{u},0,\vec{z},T)\, g(\vec{z})=
\int d^3u \int d^3z  \, m(\vec{u},\vec{z}),
\ee
where $m(\vec{u},\vec{z})=f(\vec{u}) k(\vec{u},0,\vec{z},T)\, g(\vec{z})$, and we impose the (Schr\"{o}dinger's) boundary data restriction
\be
\begin{split}
\int d^3 u \, m(\vec{u},\vec{z})= \rho (\vec{z},T),\\
\int d^3 z  \, m(\vec{u},\vec{z}) = \rho (\vec{u},0).\\
\end{split}
\ee

It is known, that once a   suitable integral kernel $k(\vec{y},s,\vec{x},t)$ is selected, then a unique solution
of the boundary data problem, in terms of functions $f(\vec{y})$ and $g(\vec{x})$,   can be obtained, \cite{zambrini1,zambrini2,olk,gar5,schr,pavon,zambrini3}.

Accordingly, $\rho  = \theta^* \theta $, provides a probability density function, which interpolates between the boundary pdfs in the time interval $[0,T]$.  The  (forward)   transition probability density function of the  inferred  Markovian diffusion process, is given in the form
\be
p(\vec{y},s,\vec{x},t) = k(\vec{y},s,\vec{x},t){\frac{\theta (\vec{x},t)}{\theta (\vec{y},s)}}
\ee
and clearly gives rise to $\rho (\vec{x},t)= \int p(\vec{y},s,\vec{x},t) \rho _B(\vec{y},s) d^3y$.

We can also introduce another (backward) transition probability density function
\be
p^*(\vec{y},s,\vec{x},t) =    k(\vec{y},s,\vec{x},t){\frac{\theta ^* (\vec{y},s)}{\theta ^*(\vec{x},t)}}
\ee
which induces a backward in time evolution of $\rho (\vec{x},t)$. Indeed,  $\int  p^*(\vec{y},s,\vec{x},t)\rho_B (\vec{x},t) d^3x  = \rho (\vec {y},s)$, and allows to infer the backward drift $\vec{b}_*$ of the diffusion process.

We point out that standard approaches to the Schr\"{o}dinger interpolation problem involve contractive semigroups, whose generators are Hermitian, \cite{zambrini1,zambrini2,olk,blanch}. This is however not a must, since non-Hermitian setting may do a job as well, \cite{zambrini}.

Given a transition probability density function $p(\vec{y},s,\vec{x},t)$   of the  (Markovian) diffusion process
\be
\rho (\vec{x},t)= \int p(\vec{y},s,\vec{x},t) \rho (\vec{y},s) d^3y.
\ee
 The factorised form of  $\rho = \theta^* \theta $,  Eq.  (108), is here admissible as well.

 We demand $\rho (\vec{x},t)$ to obey the Fokker-Plack equation in the standard form (2),
 \be
 \partial _t \rho = (1/2)\Delta  \rho - \vec{\nabla }(\vec{b} \rho ) = - \vec{\nabla}\cdot (\vec{v} \rho ).
 \ee
We recall that $\vec{v} = \vec{b} - {\frac{1}2} \vec{\nabla } \ln \rho $, and we have   other  useful  relations:  $\vec{b}_*= \vec{b} - \vec{u}$, $\vec{u}= \vec{\nabla } \ln \rho $.

  The forward drift  $\vec{b}$ derives from the general stochastic (Ito) formula,  universally  valid for Markovian diffusion processes, \cite{nelson0,nelson,zambrini1,zambrini2}
 \be
 D \vec{X}(t) = \vec{b}(\vec{x},t)=  \lim_{\Delta t \downarrow 0}  {\frac{1}{\Delta t}}
 \int (\vec{y}- \vec{x}) p(\vec{x},t,\vec{y}, t+\Delta t) d^3y .
\ee

The definition (115) of the  forward  drift field for a Markovian diffusion process, actually is a special case of the more general (Ito) formula,   valid for any smooth function $f(\vec{x},t)$ of the random variable $\vec{X}(t)$, \cite{nelson0,nelson,zambrini2}, see also \cite{morato}:
\be
\begin{split}
\lim_{\Delta s \downarrow 0}  {\frac{1}{\Delta s}}  \left[ \int p(\vec{x},t,\vec{y},t +\Delta t)
f(\vec{y},t + \Delta t)d^3y - f(\vec{x},t) \right] \\
= D f(\vec{X}(t),t)= \left[\partial _t + (\vec{b}\cdot \vec{\nabla })  + {\frac{1}2} \Delta \right] f(\vec{x},t)\\
\end{split},
\ee
where $\vec{X}(t)\equiv  \vec{x}$.

The  forward drift formally   appears through  $(D (\vec{X})(t)= \vec{b}(\vec{x},t)$, $\vec{X}(t)= \vec{x}$.  The  forward acceleration field is introduced accordingly, \cite{zambrini1,zambrini2}:
\be
D^2 \vec{X}(t) = D \vec{b}(\vec{X}(t),t) = \partial _t \vec{b} + (\vec{b} \cdot \vec{\nabla })\vec{b} + {\frac{1}2} \Delta \vec{b}.
\ee

 We can as well evaluate the backward drift
 \be
 D_*\vec{X}(t)= \vec{b}_*(\vec{x},t)=\lim _{\Delta t \downarrow 0}  \int  (\vec{x} - \vec{y}) p_*(\vec{y}, t- \Delta t, \vec{x}, t) d^3x
 \ee
 and the backward acceleration formula, \cite{zambrini,zambrini2}
 \be
 D^2_* \vec{X}(t) = D_* \vec{b}_*(\vec{X}(t))=
 \partial _t \vec{b}_*  + (\vec{b}_* \cdot \vec{\nabla })\vec{b}_* - {\frac{1}2} \Delta \vec{b}_* .
\ee

\subsubsection{Gradient case: $\vec{V} = \vec{\nabla S}$}

In conformity with  (88), let us denote
\be
\begin{split}
\vec{b} =\vec{V} + \vec{U} = \vec{\nabla }(\overline{S} + \overline{R}),\\
\vec{b}_*= \vec{V} - \vec{U} = \vec{\nabla } (\overline{S} - \overline{R}).\\
\end{split}
\ee
We get, \cite{zambrini2}:
\be
D^2 \vec{X}(t) = D (\vec{V} + \vec{U})= \partial_t(\vec{V}  + \vec{U})  + (\vec{V} + \vec{U})\cdot \vec{\nabla }\vec{V}   + {\frac{1}2} \Delta (\vec{V} + \vec{U})  + (\vec{V} + \vec{U})\cdot \vec{\nabla }\vec{U}
\ee
and, analogously
\be
D_*^2 \vec{X}(t) = D_* (\vec{V} - \vec{U})= \partial_t(\vec{V}  - \vec{U})  + (\vec{V} - \vec{U})\cdot \vec{\nabla }\vec{V}   - {\frac{1}2} \Delta (\vec{V} - \vec{U})  + (\vec{V} - \vec{U})\cdot \vec{\nabla } \vec{U}.
\ee

The stochastic version of the second Newton law presently reads:
\be
(D^2 + D_*^2) \vec{X}(t)= \partial _t\vec{V} + \vec{V}\cdot \vec{\nabla }\vec{V}  +  \vec{\nabla }\overline{Q},
\ee
where $\overline{Q}$ stems from Eq. (94), and we remember that  $\vec{U}$ is a gradient field, $\vec{U}=\vec{\nabla }\ln \overline{\rho }$.
In view of the second equation (93), we thus get
\be
(D^2 + D_*^2) \vec{X}(t) = \vec{\nabla }\Omega ,
\ee
which is   a diffusion-induced analog of the second Newton law.

\subsubsection{Non-gradient case: $\vec{V} = \vec{\nabla } \overline{S} + \vec {A} $. }

Presently we need to recast the gradient definition (120) of forward and backward drifts, in conformity with
the assumption  (97), whose Euclidean form is $\vec{V} = \vec{\nabla } \overline{S} + \vec {A} $.

Essentially nothing changes in the reasoning  (114)-118). However, presently (119) takes the non-gradient form:
\be
\begin{split}
\vec{b} =\vec{V} + \vec{U} = \vec{\nabla }(\overline{S} +   \overline{R})+ \vec{A} ,\\
\vec{b}_*= \vec{V} - \vec{U} = \vec{\nabla } (\overline{S}  - \overline{R}) + \vec{A}.\\
\end{split}
\ee
We notice that the formulas (121) and (122) retain their validity, and likewise the formula (123).

However now, in view of the non-gradient form of the current velocity field,  we need to refer to equations (102) and (103). Since (103) rewrites as
\be
\partial_t \vec{V}  + (\vec{V} \cdot \vec{\nabla })\vec{V}   +  \vec{\nabla } \overline{Q} =  - {\cal{F}}_{Lorentz},
\ee
we recover  the stochastic second Newton law in the form
\be
(D^2 + D^2_*)\vec{X}(t) =  -  {\cal{F}}_{Lorentz}
\ee
with the (anticipated) sign inversion of the Lorentz force (c.f. for comparison (100)), which is a signature of    the Euclidean version of the second Newton law.

\subsubsection{Resume.}

The above discussion effectively  sets link with a family of   diffusion processes, which  stems from  the  stochastic differential equation  (of the infinitesimal form (1)) with the forward drift
\be
 \vec{F}(\vec{x},t)= {\frac{\vec{\nabla } \theta (\vec{x},t)}{\theta (\vec{x},t)}}  + \vec{A}(\vec{x}(t)).
\ee
Notice, that by skipping the magnetic input $\vec{A}$, we are  left with the well developed framework  of conditioned diffusopn processes, \cite{zambrini1,zambrini2,gar},  within which the Hermitian  motion generator  $H_{st}$, Eq. (38) might appear as a generator of the Schr\"{o}dinger semigroup, whose kernel is defined via the Feynman-Kac formula.

With the corresponding  transition probability  density  (111) and (112)  in hands we can evaluate the
 backward drift  $ \vec{F}_*(\vec{x},t)= - \vec{\nabla } \ln \theta ^*  (\vec{x},t) +   \vec{A}(\vec{x})$.
The current velocity  (now we  turn back to the notation of sections I to V) reads    $
 \vec{v}=  {\frac{1}2}  \vec{\nabla } \ln {\frac{\theta }{\theta _*}}  + \vec{A}$.

The  factorised  probability density $\rho  = \theta _* \theta $ and the current velocity $\vec{v}$  obey  two coupled  equations (the Fokker-Planck   equation  is hereby transformed into a continuity equation):
  \be
  \begin{split}
  \partial _t \rho =  - \vec{\nabla } (\rho \vec{v}), \\
 \partial _t \vec{v} + (\vec{v}\cdot \vec{\nabla }) \vec{v}   = - [{\cal{F}}_{Lorentz}  + \vec{\nabla } Q],     \\
  \end{split}
   \ee
c.f. Eqs. (101)  and  (103).

  We may  cross-check  the validity of  formulas for $\vec{F}$, $\vec{F}^*$  and  $v$, by taking $\vec{F}$ for granted and resorting to the rewriting  of the  F-P equation as the continuity equation.    Namely, we have $\partial _t \rho = - \vec{\nabla }\cdot [(\vec{F} - \vec{\nabla }\ln \rho ^{1/2})\rho (\vec{x},t)]$.  Inserting $\vec{F}=\vec{\nabla }\ln \theta + \vec{A}$, and $\rho = \theta_* \theta $, we  readily arrive at  $\partial _t \rho =  - \vec{\nabla } (\rho \vec{v})$ with  $\vec{v}$ given  above.

 Analogously,  by means of direct substitutions, we may check that   $\partial _t \rho =  - \vec{\nabla } (\vec{v} \rho )$ can be rewritten as the backward Fokker-Planck equation $-\partial _t \rho = {\frac{1}2} \Delta \rho + \vec{\nabla }\cdot (\vec{F}^* \rho )$, $t \in [0,T]$, \cite{gar,nelson}. We note the validity of the formula $\vec{F}^* = \vec{F} - \vec{\nabla } \ln \rho $. \\

This completes the identification of "magnetic perturbations" in the present framework, while  allowing to bypass   obstacles identified in Section VI, by admitting  perturbations of  Euclidean generators $H_{Eucl}$ and $H_{Eucl}^*$ by   suitable  potentials denoted  $\Omega $  in the above

\section{Outlook}

Although motivated by $d\geq 3$ considerations of Ref. \cite{monthus}, we took a liberty to stay on the level of $R^3$, where the  potential novelties  of the  non-Hermitian  theoretical framework  could have been thoroughly tested against the current status of what is named  the {\it Brownian motion in a magnetic field}, \cite{czopnik,aquino,aquino1,abdoli}.  An additional impetus came from the need to reconsider some observations  of Ref. \cite{gar}, where the probabilistic status of the integral kernel of $\exp(-tH_{Eucl})$ has been questioned.
The probabilistic significance issue  for integral kernels generated by  non-Hermitian operators   has proved to be vital for the understanding of non-conservative   diffusion processes, and specifically their links (affinities, analogies, similarities) with electromagnetic perturbations of diffusing particles.

The  related  pros and cons discussion has been carried out logically.  In Section II we gave an outline  of the theoretical framework of \cite{monthus}, while restricted to $R^3$.
Section III has been devoted to"magnetic"-looking Lagrangian dynamics, which actually is a Euclidean version of the standard (classic) one,  with an  emphasis on Lorentz force connotations. Our  main motivation was  to  deduce  (most useful in the quadratic case)  Lagrangian action functionals,  and next  complete a standard route towards an evaluation of path integral expressions for involved propagators (actually transition pdfs, with  reference to diffusion processes).

In Section IV we have recalled the original appearance of magnetic perturbations in the standard picture of the  Brownian motion, \cite{czopnik}, followed  in Section V by a transcription to spatial  non-conservative processes, and paid some attention to a possible meaning of retained magnetic perturbation features (not completely lost in "translation").

In view of the  significance of   $H_{Eucl}$  in  the discussion of  Ref. \cite{monthus},   we have performed   in Section V a complete evaluation of the integral kernel of $\exp(-tH_{Eucl})$, by starting  from the first path integration   principles. We have identified a numer of defective properties of this kernel, which strongly limit its probabilistic significance.  For  probabilistic applications  (e.g. construction of transition probability densities of conditioned stochastic processes,  \cite{olk,zambrini})  we need  strictly positive kernel functions, satisfying the semigroup composition law.

These properties were employed in section VII, where   we have proposed a remedy to obstacles induced by the "bare" $H_{Eucl}$ generator of motion.  This amounts to an alternative  view  (if compared with  Brownian motion standards, \cite{czopnik})  on the "magnetic" dynamics of non-conservative diffusion processes. Eqs.  (126)-(128)   succinctly  summarize  this endeavour,  by making explicit the  (electro)"magnetic" impact  on diffusion currents (strictly speaking, on the deduced current velocity of the diffusion process).

We find worth mentioning  our   attention paid to the identification   of diffusion currents, and specifically  to  the  dynamics of currents and   related  current velocities.

\begin{appendix}

\section{Evaluation of   $\exp[-t H_{Eucl}] (\vec{x}, \vec{y})$.}

\subsection{The Lagrangian.}

The integral kernel $k(\vec{y},0,\vec{x},t)=\exp(-t H_{Eucl})(\vec{x},\vec{y})$, where $H_{Eucl} $, in case of quadratic Lagrangians is known to be  proportional to $\exp{(-S)}$,  where $S$ stands for the classical action functional. The  normalization   factor  is   given by the van Vleck formula, \cite{monthus,grosche},  so that  $k(\vec{y},0,\vec{x},t)$ takes the form
\be
k(\vec{y},0,\vec{x},t) = {\frac{1}{2\pi }^{3/2}}   \sqrt{\det \left(- {\frac{\partial^2 S(\vec{y},0,\vec{x},t)}{\partial_{x_i} \partial_{y_j}}}\right)}  e^{ - S(\vec{y},0,\vec{x},t)}.
\ee

Recalling (14) and (27), we  specify the Lagrangian   ${\cal{L}}_{Eucl}= {\frac{1}2} \dot{\vec{x}}^2 - \dot{\vec{x}} \cdot \vec{A}$. The  solenoidal vector field has the form $\vec{A}= (-y,x,0)$, so that ${\frac{1}2}  \vec{\nabla } \times \vec{A})=(0.0.1)$.

From now on,  we skip  the Euclidean label in $\mathcal{L}_{Eucl}$, since all arguments refer to the Euclidean setting.
 The Lagrangian, to be employed in the path integral procedure (14), (15), has the form:

\be
\mathcal{L}(x,y,\dot{x},\dot{y},\dot{z})=\frac{1}{2}\left(\dot{x}^2+\dot{y}^2+\dot{z}^2\right)-(\dot{x} y-x\dot{y})=\mathcal{L}_1(x,y,\dot{x},\dot{y})+\mathcal{L}_2(\dot{z}),
\ee
where
\be
\mathcal{L}_1(x,y,\dot{x},\dot{y})=\frac{1}{2}\left(\dot{x}^2+\dot{y}^2\right)-(\dot{x} y-x\dot{y})
\ee
and
\be
\mathcal{L}_2(\dot{z})=\frac{1}{2}\dot{z}^2.
\ee

\subsection{Evaluation of the action functional.}

Let $S$ be the Euclidean  action
\be
S=\int\limits_0^t\mathcal{L}(x,y,\dot{x},\dot{y},\dot{z})ds=\int\limits_0^t\mathcal{L}_1(x,y,\dot{x},\dot{y})ds+
\int\limits_0^t\mathcal{L}_2(\dot{z})ds=S_1+S_2.
\ee
The Euler-Lagrange equations are of the form
\be
\begin{split}
&\frac{\partial\mathcal{L}}{\partial x}-\frac{d}{dt}\frac{\partial\mathcal{L} }{\partial \dot{x}}=\dot{y}-\frac{d}{dt}(\dot{x}-y)=2\dot{y}-\ddot{x}=0,\\
&\frac{\partial\mathcal{L}}{\partial y}-\frac{d}{dt}\frac{\partial\mathcal{L} }{\partial \dot{y}}=-\dot{x}-\frac{d}{dt}(\dot{y}+x)=-2\dot{x}-\ddot{y}=0,\\
&\frac{\partial\mathcal{L}}{\partial z}-\frac{d}{dt}\frac{\partial\mathcal{L} }{\partial \dot{z}}=-\frac{d}{dt}\dot{z}=0.
\end{split}
\ee
Let the initial conditions have the form $(x(0),y(0),z(0))=(x_1,x_2,x_3)$ and $(x(t),y(t),z(t))=(y_1,y_2,y_3)$. From the last equation of the system (79) we get
\be
z(s)=c_1+c_2 s,
\ee
so after taking into account the boundary conditions
\be
z(0)=c_1=x_3,\qquad z(t)=c_1+c_2 t=y_3,
\ee
we obtain
\be
z(s)=x_3+\frac{y_3-x_3}{t} s.
\ee

The action $S_2$ can be readily calculated:
\be
S_2=\int\limits_0^t \frac{1}{2}\dot{z}^2\,ds=\frac{1}{2}\int\limits_0^t\left(\frac{y_3-x_3}{t}\right)^2\,ds=\frac{(y_3-x_3)^2}{2t}.
\ee
The first two equations of the system of equations (A.8) are slightly more complicated.
From the first equation of (A.8) we get $\dot{y}=\ddot{x}/2$,  which upon substituting to
 the second equation, leads to
\be
4\dot{x}+\dddot{x}=0.
\ee
The solution of this equation reads
\be
x(s)=c_1+c_2\cos(2s)+c_3\sin(2s).
\ee
By substituting the obtained  $x(s)$ to $\dot{y}=\ddot{x}/2$ we get
\be
\dot{y}(s)=-2c_2\cos(2s)-2c_3\sin(2s),\qquad y(s)=c_4-c_2\sin(2s)+c_3\cos(2s).
\ee

Given the initial conditions, we obtain the following system of  equations, from which
$c_i, i=1,2,3,4$ need to be retrieved:
\be
\begin{split}
x(0)&=c_1+c_2 =x_1,\\
x(t)&=c_1+c_2\cos(2t)+c_3\sin(2t)=y_1,\\
y(0)&=c_4+c_3= x_2,\\
y(t)&=c_4-c_2\sin(2t)+c_3\cos(2t)=y_2.
\end{split}
\ee
Its solution reads
\be
\begin{split}
c_1&=\frac{1}{2}\left(x_1+y_1+(y_2-x_2)\cot t\right),\\
c_2&=\frac{1}{2}\left(x_1-y_1-(y_2-x_2)\cot t\right),\\
c_3&=\frac{1}{2}\left(x_2-y_2+(y_1-x_1)\cot t\right),\\
c_4&=\frac{1}{2}\left(x_2+y_2+(y_1-x_1)\cot t\right).
\end{split}
\ee

After substituting (A.16), (A.17) to solutions (A.14), (A.15),  and next performing integrations,  we can evaluate $S_1$:
\be
S_1=\int\limits_0^t\left[\frac{1}{2}\left(\dot{x}^2+\dot{y}^2\right)-(\dot{x} y-x\dot{y})\right]\,ds=x_1y_2-x_2y_1+\frac{1}{2}\left[(y_1-x_1)^2+(y_2-x_2)^2\right]\cot t .
\ee
Accordingly, the  classical action functional reads
\be
S=S_1+S_2=x_1y_2-x_2y_1+\frac{1}{2}\left[(y_1-x_1)^2+(y_2-x_2)^2\right]\cot t+\frac{(y_3-x_3)^2}{2t} .
\ee

\subsection{Evaluation of the van Vleck (Morette-van Hove) factor.}

Let  the action functional  $S$ be given by Eq. (A.19). Then, the  van Vleck matrix   reads
\be
-\frac{\partial^2 S}{\partial x_n\partial y_m}=
\left( {\begin{array}{ccc}
		\cot t & 1 & 0 \\
		-1 & \cot t & 0 \\
		0 & 0 & \frac{1}{t} \\
\end{array} } \right),\label{r2}
\ee
where $n, m = 1,2,3$. The determinant of this matrix   reads
\be
\det\left(-\frac{\partial^2 S}{\partial x_n\partial y_m}\right)=\frac{1}{t\sin^2 t}.\label{r3}.
\ee
Hence, the normalizing factor takes the form  (the square root is taken as an arithmetic one in $R^+$
\be
\sqrt{\frac{\det\left(-\frac{\partial^2 S}{\partial x_n\partial y_m}\right)}{(2\pi)^3}}=\frac{1}{2\pi |\sin t|}\left(\frac{1}{2\pi t}\right)^{1/2}.\label{r4}
\ee
This gives us the final form of the  sought for  propagator
\be
k(\vec{y},0,\vec{x},t)=\frac{1}{2\pi |\sin t|} \left(\frac{1}{2\pi t}\right)^{1/2}\exp\left\{-x_1y_2+x_2y_1-\frac{1}{2}\left[(y_1-x_1)^2+(y_2-x_2)^2\right]\cot t-\frac{(y_3-x_3)^2}{2t}\right\}.
\ee

We note, that the functional form of $\exp[-t H^*_{Eucl}] (\vec{x}, \vec{y})$,  differs from this
  given in Eq. (A.21), by   merely  changing a sign of the $x_1y_2-x_2y_1$ contribution in the exponent. This corresponds to the  sign inversion  of the involved vector potential $\vec{A}$.

\subsection{Verification of the semigroup composition law.}

If we turn back to the diffusion process (1)-(3),  the semigroup  identity  $\exp(sL^*) \exp[(t-s)L^*] = \exp(-tL^*)$,  is related to the Chapman-Kolmogorov  property of transition pdfs:  $p(\vec{z},0,\vec{x},t)=\int p(\vec{z},0,\vec{y},s) p(\vec{y},s,\vec{x},t)d^3y$, with $0<s<t$.
It is by no means obvious, that after  passing from $- L^*= H_{Eucl} + {\cal{V}}$ to the "bare"  generator  $H_{Euclid}$,  the   integral kernel   of $\exp(-tH_{Eucl})$  inherits the previous composition  property: $p_s p_{t-s} = p_t \Rightarrow k_s k_{(t-s)} = k_t$.

Presuming that the  kernel  function (A.24) is a valid   integral  kernel of $\exp(-tH_{Eucl})$,  we  shall explicitly  verify  the validity of the composition rule    $\exp(-sH_{Eucl}) \exp[-(t-s)H_{Eucl}] = \exp(-tH_{Eucl})$, directly    in terms of integral kernels.  This amounts to  demanding :
\be
k(\vec{z},0,\vec{x},t)=\int k(\vec{z},0,\vec{y},s)k(\vec{y},s,\vec{x},t)d^3y,\label{r6}
\ee
where  the integration is over $R^3$.

The right-hand side of the above equation is of the form
\be
\begin{split}
&\int k(\vec{z},0,\vec{y},s)k(\vec{y},s,\vec{x},t)d^3y=
\frac{1}{(2\pi)^3}\frac{1}{\sqrt{s(t-s)}|\sin s\sin(t-s)|}\\
& \int \exp\left\{  -x_1y_2+x_2y_1-\frac{1}{2}\left[(y_1-x_1)^2+(y_2-x_2)^2\right]\cot s-\frac{(y_3-x_3)^2}{2s}\right.\\
& \left. -y_1 z_2+y_2z_1-\frac{1}{2}\left[(z_1-y_1)^2+(z_2-y_2)^2\right]\cot (t-s)-\frac{(z_3-y_3)^2}{2(t-s)}\right\}dy_1dy_2dy_3.\label{r7}
\end{split}
\ee

One must be careful, while evaluating the Gaussian  integrals.  A single $y_3$ integration is unproblematic, since
\be
I_3 = \int dy_3 \exp\left\{-\frac{(y_3-x_3)^2}{2s}-\frac{(z_3-y_3)^2}{2(t-s)}\right\}= \sqrt{\frac{2\pi s(t-s)}{t}}\exp\left[-\frac{(z_3-x_3)^2}{2t}\right].\label{r8}
\ee
The trouble appears,  if  we pass to the  integration over $y_2$. We are working with   Gaussian integrals, hence  factors facing $ y_2^2$  in (A.26) should be negative.   For instance,  if $s\in(\pi/2,\pi)$ and $t-s\in(\pi/2,\pi)$ the integration with respect to  $y_2$, along the whole real line $R$, is impossible.

In case, when the  integration over $y_2$   produces a finite outcome  (e.g. when $cot s$ and $cot(t-s)$ are positive),  we obtain
\be
\begin{split}
I_2 &= \int dy_2 \exp\left\{ -x_1y_2+y_2z_1 -\frac{1}{2}(y_2-x_2)^2\cot s-\frac{1}{2}(z_2-y_2)^2\cot (t-s)\right\}\\
&=\sqrt{\frac{2\pi \sin s\sin (t-s)}{\sin t}}\exp\left\{\frac{1}{2}\left[-z_2^2\cot (t-s)-x_2^2\cot s +\frac{(z_1-x_1+z_2\cot (t-s)+x_2\cot s)^2}{\cot (t-s)+ \cot s}\right]\right\}.\label{r9}
\end{split}
\ee
The integration  with respect to  $y_1$ involves the same precautions as in the  case of $y_2$, leading to
\be
\begin{split}
I_1 &= \int dy_1 \exp\left\{ -y_1z_2+x_2y_1 -\frac{1}{2}(y_1-x_1)^2\cot s-\frac{1}{2}(y_1-z_1)^2\cot (t-s)\right\}\\
&=\sqrt{\frac{2\pi \sin s\sin (t-s)}{\sin t}}\exp\left\{\frac{1}{2}\left[-z_1^2\cot (t-s)-x_1^2\cot s +\frac{(x_2-z_2+z_1\cot (t-s)+x_1\cot s)^2}{\cot (t-s)+ \cot s}\right]\right\}.\label{r10}
\end{split}
\ee

An ultimate outcome, clearly is:
\be
\begin{split}
&\int k(\vec{z},0,\vec{y},s)k(\vec{y},s,\vec{x},t)d^3y=
\frac{1}{(2\pi)^3}\frac{1}{\sqrt{s(t-s)}|\sin s\sin(t-s)|}I_1\cdot I_2\cdot I_3=\\
&\frac{1}{(2\pi)^{3/2}}\frac{1}{\sqrt{t}}\frac{1}{|\sin t|}\exp\left\{-x_1z_2+x_2z_1-\frac{1}{2}\left[(z_1-x_1)^2+(z_2-x_2)^2\right]\cot t-\frac{(z_3-x_3)^2}{2t}\right\}.\label{r11}
\end{split}
\ee

The composition rule surely leads to $k(\vec{z},0, \vec{x},t)$ of the form (A.21),   for   all  $0<s<t<\pi/2$,
when  a simultaneous  positivity  of $\sin s$, $\sin t$,  $\sin (t-s)$, and $\cot t $  is secured.  Notice that the $\sin t$ positivity extends to $0<s<t< \pi $, but  for $t \in (\pi/2, \pi )$ the coefficient $\cot t$ in the   gaussian exponent becomes negative and explodes to $-\infty$ as $t\to \pi $. This rules out a  consistent  continuous extension of the   $t\in (0, \pi /2) $    time domain,  except for a sequence of  disjoint  time intervals $t\in (2n\pi,   2n\pi + \pi/2)$, with $n=0,1,2,...$.

\end{appendix}

\end{document}